\documentclass[preprint]{aastex}	

\newcommand{\ergs}{ergs~s$^{-1}~$}
\newcommand{\mdot}{$\dot{{M}}$ } 

\received{}
\accepted{}
\lefthead{Bloser et al.}
\righthead{Spectral variation in 4U 1820-30 with {\em RXTE}}

\begin{document}

\title{{\em RXTE} Studies of Long-Term X-ray Spectral Variations\\ 
in 4U 1820-30}
\author{P. F. Bloser, J. E. Grindlay, P. Kaaret}
\affil{Harvard-Smithsonian Center for Astrophysics, \\
60 Garden Street, Cambridge, MA 02138 \\
pbloser@cfa.harvard.edu}
\author{W. Zhang, A. P. Smale}
\affil{Laboratory for High Energy Astrophysics,
Goddard Space Flight Center, \\Greenbelt, MD 20771\\
zhang@xancus10.gsfc.nasa.gov}
\and
\author{D. Barret}
\affil{Centre d'Etude Spatiale des Rayonnements, CNRS/UPS,\\
9 Avenue du Colonel Roche, 31028 Toulouse Cedex 04, France\\
Didier.Barret@cesr.fr}


\begin{abstract}
We present the results of detailed spectral studies of the 
ultra-compact low mass X-ray binary (LMXB) 4U 1820-30 carried out 
with the {\em Rossi X-ray Timing Explorer (RXTE)} during 1996-7. 4U 1820-30
is an ``atoll'' source X-ray burster (XRB) 
located in the globular cluster NGC 6624. It is known to 
have an 11 minute binary period and a $\sim 176$ day 
modulation in its 2--12 keV flux. Observations were made with 
the PCA and HEXTE instruments on {\em RXTE} at roughly one-month 
intervals to sample this long-term period and study 
flux-related spectral changes. We find that the 176-day period
corresponds to normal atoll source motion in the color-color diagram,
indicating that the long period is due to mass accretion rate
changes.  There are clear correlations 
between our fitted spectral parameters and both the broad-band (2--50 keV) 
flux and the position in the color-color diagram, as described by the
parameter $S_a$ introduced by M\'endez et al. (1999).  The hard X-ray
tail becomes more prominent at  
low flux levels as the source moves into the island state. In
addition, we find a strong correlation
between the position in the color-color diagram and the frequencies of
the kilohertz quasi-periodic 
oscillations (kHz QPOs)
reported by Zhang et al. (1998). This lends further support to the 
notion that evidence for the last stable orbit in the 
accretion disk of 4U 1820-30 has been observed.  For a model
consisting of Comptonization of cool photons by hot electrons plus an
additional blackbody component, we report an abrupt change in the spectral
parameters at the same accretion rate at which the kHz QPOs disappear,
suggesting a change in the accretion flow that affects both the
spectrum and the timing properties.  For a model consisting of a
multicolor disk blackbody plus a cut-off power law, we find that the
inner disk 
radius reaches a minimum at the same accretion rate at which the kHz QPO
frequency saturates, as expected if the disk reaches the last stable
orbit.  Both models face theoretical and observational 
problems when interpreted physically for this system.
\end{abstract}
\keywords{accretion, accretion disks --- stars: individual (4U
1820-30) --- stars:neutron --- X-rays:stars}

\section{Introduction}
\label{sec-intro}

The broad-band spectral sensitivities of two X-ray
astronomy spacecraft, the {\em Rossi X-ray Timing Explorer} ({\em
RXTE}, Bradt, Rothschild, \& Swank 1993) and {\em BeppoSAX} (Boella et
al. 1997), combined with the high timing resolution of {\em RXTE},
have allowed major advances in recent years in the study of Low Mass
X-ray Binaries (LMXBs).  Progress has been particularly rapid in the
field of X-ray bursters (XRBs), binary systems containing a
low-magnetic field neutron star (NS) accreting matter from a late-type
companion star via Roche lobe overflow.  It had become apparent by the
mid 1990's that NS systems were capable of producing hard X-ray ($\gtrsim
20$ keV) emission, as indicated by BATSE and SIGMA detections of
several XRB at $\sim 100$ keV (Barret \& Vedrenne 1994, Tavani \&
Barret 1997 and references therein).  This eliminated hard X-ray
radiation as a unique identifier of black hole (BH) systems, but the limited
sensitivity and spectral resolution of these instruments, together
with the lack
of simultaneous soft X-ray observations, did not allow the spectral
shape to be studied in detail or possible differences between NS and
BH hard X-ray properties to be determined.  The two
new satellites are revolutionary due to their broad spectral coverage:
the Proportional Counter Array (PCA) and High Energy X-ray Timing
Experiment (HEXTE) instruments on {\em RXTE} cover the
range 2--150 keV, while the suite of narrow-field instruments (NFIs)
on {\em BeppoSAX} 
combine to cover 0.1--200 keV.  Many XRB have now been observed with
these experiments, allowing detailed modeling of the complete X-ray to
hard X-ray spectra with a variety of complex models (e.g., Church et
al. 1998; Guainazzi et al. 1998; in 't Zand et al. 1999; Olive et
al. 1999, Piraino et al. 1999a; Piraino et al. 1999b; Barret et
al. 1999, 2000).

In addition to broad spectral coverage, {\em RXTE} offers
unprecedented timing resolution ($\sim 1 \mu$s) due to the large
collecting area of 
the PCA.  This has led directly to the discovery of quasi-periodic
oscillations (QPOs) with frequencies near 1000 Hz (kHz QPOs) in the
X-ray timing power spectra of many NS systems (Strohmayer et al. 1996;
van der Klis et al. 1996).  Since kilohertz frequencies are naturally
associated with orbital timescales near a NS, it is generally accepted
that the kHz QPOs are produced in the inner accretion disk and that
the frequency is related to the inner disk radius.  If this is true, the kHz
QPOs are a potentially valuable probe of strong field gravity (Kaaret
\& Ford 1997).  In particular, the relationship between QPO frequency
and accretion rate should provide a means to test for the existence of
the marginally stable orbit predicted by general relativity (Kaaret,
Ford, \& Chen 1997; Zhang, Strohmayer, \& Swank 1997).

In this paper we present the results of {\em RXTE} observations of 4U
1820-30, an
XRB with a particularly colorful history.  4U 1820-30 is located in
the core of
the globular cluster NGC 6624 and was one of the first globular
cluster X-ray sources discovered (Giacconi et al. 1974).  The distance
to NGC 6624, $6.4 \pm 0.6$ kpc, was derived by Vacca et al. (1986)
from a study of horizontal branch stars.  Thus the distance to 4U
1820-30 is among the better known of all X-ray sources.  It is the
brightest globular cluster source, and was the first identified X-ray
burster (Grindlay et al. 1976).  The source has been observed to shift
between a high and a low state with a period of about 176 days
(Priedhorsky \& Terrell 1984); in the All Sky Monitor (ASM) on {\em
RXTE} the 2--12 keV flux varies by typically a factor of $\sim 3$ and
as much as a factor of $\sim 6$ between the two
states (see Figure~\ref{fig-asm}).  Also observed is an 11 minute (685
s) period believed to be the 
orbital period, the shortest known orbital period of all cosmic binary
systems (Stella, Priedhorsky, \& White 1987).  Such a short period
suggests that the mass-donating star is a helium dwarf with a mass of
0.06--0.08 M$_{\odot}$ (Rappaport et al. 1987).  Two simultaneous kHz
QPOs were detected in the persistent flux from 4U 1820-30 by Smale,
Zhang, \& White (1997).

The goal of the present investigation is to observe the broad-band
(2--150 keV) spectrum of 4U 1820-30 with the PCA and HEXTE instruments
on {\em RXTE} and to study the relationships between spectral
parameters, kHz QPO frequencies, and accretion rate.  The steady
176-day variation of the flux from 4U 1820-30 is most likely due to
modulation of the accretion flow, since X-ray bursts are only observed
in the low state (Stella, Priedhorsky, \& White 1987), and so this
source provides a good laboratory for the study of accretion-related
spectral changes.  We have observed 4U 1820-30 over the course of
a year at roughly monthly intervals to sample the entire 176-day
period.  This same data set was used to study timing properties by
Zhang et al. (1998).  In Section~\ref{sec-prev} we describe previous
spectral and 
timing observations of 4U 1820-30.  In Section~\ref{sec-obsanal} we
describe the observations and our analysis procedure.  The results are
presented in Section~\ref{sec-res}, discussed in
Section~\ref{sec-disc}, and we state our conclusions in
Section~\ref{sec-conc}. 

\section{Previous Observations of 4U 1820-30}
\label{sec-prev}

As evidenced by its long history, 4U 1820-30 is an extremely
well-studied system.  In this section we briefly review previous
spectral and timing observations of the source.

\subsection{Spectral Observations}
\label{sec-prevspec}

4U 1820-30 has previously been observed by {\em ANS}, {\em EXOSAT},
{\em Ginga}, 
{\em ASCA}, {\em Einstein}, and {\em BeppoSAX}.  A wide variety of
models have been used to fit the recorded spectra, but the most
successful have been two-component models consisting of either
optically thin thermal
bremsstrahlung plus a blackbody (TB + BB), a power law with exponential cutoff
(an analytical approximation of unsaturated Comptonization) plus a
blackbody (CPL + BB), or a more detailed Comptonization model developed
by Sunyaev \& Titarchuk (1980) plus a blackbody (CompST + BB, after
the {\em CompST} model in XSPEC).  
Models based on optically thin thermal bremsstrahlung are widely
believed to be 
unphysical for XRB, since the emission measures typically derived
($\sim 10^{60}$ cm$^{-3}$) imply, for small optical depths, that the
plasma cloud has a radius $\sim 10^{10}$ cm, too large to be heated to
several keV by the neutron star (White et al. 1986; Christian \& Swank
1997).  We 
therefore restrict our attention to the CPL and CompST (with or
without a BB)
model results.  The important parameters for the CPL + BB model are
the blackbody temperature $kT_{BB}$, the power law photon index
$\alpha$, and the exponential cutoff energy $E_c$.  For the CompST +
BB model, the parameters of interest are $kT_{BB}$ and the temperature
$kT_e$ and optical depth $\tau$ of the Comptonizing electron cloud,
assumed here to be spherical.  Photoelectric absorption at low
energies by the interstellar medium must be included as well, given by
the equivalent hydrogen column density $N_H$ (found using the
absorption cross sections of Morrison \& McCammon(1983)).  Finally, a
gaussian line feature 
at $\sim 6.7$ keV is often required, representing a blend of $K\alpha$
emission  
lines from helium-like ions of iron (Fe {\sc xxv} or {\sc xxvi}; White et
al. 1986; Hirano et al. 1987).
A summary of previously-reported spectral fits using
these models is given in Table~\ref{tab-prev} and shown graphically in
Figure~\ref{fig-comp}; we discuss them briefly here.

Intensity-related spectral changes were first reported for 4U 1820-30
by Parsignault \& Grindlay (1978) using {\em ANS}/HXX data and a power
law fit.  They found that the photon
index $\alpha$ varied between $\sim 2$ at high count rates and $\sim
1.4$ at low count rates, indicative of spectral hardening at low
luminosity.   More complex models could be used after the {\em EXOSAT}
ME instrument observed 4U 1820-30 between 1984 and 1985  (Stella,
White, \& Priedhorsky 1987), finding a high state luminosity (1--30
keV) of $6.0 \times 10^{37}$ \ergs  and a low state luminosity of $2.0
\times 10^{37}$ \ergs.  Fits with the CPL + BB model (Stella, White,
\& Priedhorsky 1987) gave low state values of $\alpha = 2.5$, $E_c >
30$ keV, and $kT_{BB} = 2.3$ keV, while in the high state $\alpha$
decreased to 1.7, $E_c$ decreased to $\sim 12$ keV, and $kT_{BB}$ fell
to 2.0 keV.  Note that the inclusion of an exponential cutoff has
changed the intensity-related behavior of $\alpha$ compared to the
{\em ANS} results; the softening of the spectrum is now described by
the lower cutoff energy while $\alpha$ describes the spectrum below
this.  White, Stella, \& Parmar (1988) fitted the CompST +  BB model
to the {\em EXOSAT} data and found that between the low and high
states $kT_{BB}$ decreased from 2.06 to 1.61 keV, $kT_e$ fell slightly
from 3.5 to 3.3 keV, and $\tau$ decreased from 13.2 to 11.9.  For both
models the ratio of blackbody to ``hard component'' flux $L_{BB}/L_H$
decreased from 50\% to 25\% as the luminosity increased.   In a
similar energy band, the {\em Ginga} LACs observed 4U 1820-30 at two
different intensity levels in  1987 May (Ercan et al. 1993),  but the
power law was steeper with a lower cutoff  than in the {\em EXOSAT}
spectrum (Table~\ref{tab-prev}).  The ratio $L_{BB}/L_H$ was similar
however.    

Lower energy coverage was provided by {\em ASCA}/GIS and {\em
Einstein} observations.  {\em ASCA} observed 4U 1820-30 in 1993  in
the low state and the spectrum (0.6--11 keV) was fitted with the
CompST + BB model (Smale et al. 1994).  CompST parameters  were
similar to those found by {\em  EXOSAT}, but the blackbody temperature
was substantially lower ($kT_{BB} = 0.76 \pm 0.02$ keV).  Christian \&
Swank (1997) report on an {\em Einstein} (SSS + MPC) observation
(0.5--20 keV) of 4U 1820-30 from 1978.  The luminosity was $5.5 \times
10^{37}$ 
\ergs, indicative of the high state, and yet fitting the CompST + BB model
gave paramters very similar to those from {\em ASCA}
(Table~\ref{tab-prev}).  The blackbody 
flux was  only 6\% of the CompST flux.  

The first observations of 4U 1820-30 extending above $\sim 30$ keV were
made by the NFIs (0.1--200 keV) on {\em BeppoSAX} in 1998 (Piraino et
al. 1999a; Kaaret et al. 1999).  These were the first truly broad-band
observations, though they covered only a narrow range in
luminosity (Table~\ref{tab-prev}).  A CPL + BB fit to the
0.3--40 keV spectrum gave $kT_{BB} = 0.47$ keV, $\alpha = 0.55$, and
$E_c = 4.5$ keV, while the CompST + BB model yielded $kT_{BB} =$
0.46--0.66 keV, $kT_e = 2.83$ keV, and $\tau = 13.7$.  Thus the
instruments with response below 1 keV consistently find low values of
$kT_{BB}$.  The entire range of
luminosities observed so far with all instruments has yielded very
similar values for the parameters of the CompST + BB model. 

The {\em BeppoSAX} Phoswich Detection System (PDS) failed to detect 4U
1820-30 
above 40 keV, consistent with the failure of BATSE to detect the
source (20--100 keV) over the first four years of the {\em CGRO}
mission (Bloser et al. 1996).  Thus 4U 1820-30 is not a member
of the class of XRB detected at $\sim 100$ keV.

\subsection{Kilohertz QPOs}

Two simultaneous kHz QPOs were discovered in the persistent emission of 4U
1820-30 by Smale, Zhang, \& White (1997) using {\em RXTE}/PCA data from 1996
October.  The characteristics of the 
QPOs were similar to those found in other XRB: their centroid
frequencies were correlated with the 2--60 keV PCA count rate, but the
frequency difference remained constant at $275 \pm 8$ Hz.  The
fractional rms amplitudes increased dramatically with energy.  Zhang et
al. (1998) studied monthly PCA observations of 4U 1820-30 (the same
data set reported on here; see Section~\ref{sec-obsanal}) and found
that the frequencies of the QPOs are only correlated with PCA count
rate below a critical value.  Above this, the QPO frequencies
remain constant while the count rate increases.  Saturation of QPO
frequency at high mass accretion rates is an expected signature of the
marginally stable orbit (Miller, Lamb, \& Psaltis 1998; Kaaret,
Ford, \& Chen 1997); however, M\'endez et al. (1999) have shown that PCA count
rate is not a good indicator of accretion rate.  Kaaret et al. (1999)
used more reliable estimators of mass accretion rate, broad-band flux
and hard color, to show that the frequency saturation is robust and
thus strong evidence that the inner disk in 4U 1820-30 does reach the
marginally stable orbit at high accretion rates.

\section{Observations and Analysis}
\label{sec-obsanal}

\subsection{{\em RXTE} Observations and Data Reduction}

The PCA instrument on {\em RXTE} is made up of five Xenon proportional
counter units (PCUs, numbered 0--4) sensitive from 2--60 keV with a
total area of about 6500 
cm$^2$ (Bradt, Rothschild, \& Swank 1993).  
The HEXTE instrument consists of two clusters of four NaI(Tl)/CsI(Na)
phoswich scintillation detectors sensitive from 15--200 keV with an
effective area of 1600 cm$^2$; these clusters rock on and off source
to obtain background measurements (Rothschild et al. 1998).  The ASM
is comprised of three Scanning Shadow Cameras (SSCs) with
one-dimensional slit masks and $6^{\circ} \times 90^{\circ}$ fields of
view that monitor the fluxes of bright X-ray sources several times a
day in the 2--12 keV band (Levine et al. 1996).  

Sixteen joint PCA/HEXTE pointed observations of 4U 1820-30 were
performed at roughly monthly intervals between 1997 February 9 and
1997 September 10.  In addition, seven observations made between 1996
October 26 and 31 were included (Smale et al. 1997).  
The {\em RXTE}/ASM 2--12 keV light curve of 4U 1820-30 during this
period is shown in Figure~\ref{fig-asm}, with the times of our pointed
observations marked.  Each point represents the one-day average of the
ASM count rate.  The 176-day modulation is clearly visible and is
well-sampled by the pointed observations.  The ASM count rate varies
from $\sim 5$ to $\sim 30$ counts s$^{-1}$ during this time, or $\sim
67$ mCrab to $\sim 400$ mCrab.  Assuming a Crab-like spectrum and $N_H
= 0.3 \times 10^{22}$ cm$^{-2}$ (see below), these 
low and high state values correspond to 1--20 keV luminosities of
$\sim 1.2
\times 10^{37}$ \ergs and $\sim 7.0 \times 10^{37}$ \ergs, respectively,
consistent with the range of luminosities observed previously
(Table~\ref{tab-prev}).  The dates, starting times in seconds of Mission
Elapsed Time (MET = seconds since 1994 Jan 1,
0h0m0s UTC), and mean PCA count
rates (2--16 keV) for all 23 observations are listed in Table~\ref{tab-log}.

The data were divided into the same 90 segments (persistent emission
only, typical length $\sim
3000$ seconds) used by Zhang et al. (1998) and reduced using the
standard {\em RXTE} analysis tools contained in FTOOLS 4.2.  For our
color-color diagram and 
PCA spectral analysis we used the ``Standard 2'' data, which provide
128 energy channels between 2 and 100 keV with 16-second time
resolution.  Spectra were extracted with {\em saextrct} v4.0b.
Backgrounds were estimated with {\em pcabackest} v2.1b 
(October 1998), and the v2.2.1 response matrices (January 1998) were used.
For the HEXTE spectral analysis we used event mode (E\_8us\_256\_DX1F)
data with the 1997 March response matrices.  Event mode data were used
instead of the standard archive mode data because the 1996 archive
data files were corrupted; this problem has been corrected in the
reprocessed distribution of 1996 data.
Spectra were extracted
with {\em saextrct} and rebinned by a factor of two using {\em grppha}
to provide the same energy bins as standard archive data below 60 keV.
Deadtime corrections were 
computed using {\em hxtdead} v0.0.1.

\subsection{Color-Color Diagram and Parameterization of Accretion Rate}

A color-color diagram (CCD) was produced from the PCA data with one
point for each of the 90 data segments.  Only PCUs 0, 1, and 2 were on
continuously during all observations, and so only data from these PCUs are
included here.
The soft and hard colors are defined as the ratios of
background-subtracted PCA count rates in the bands 
3.5--6.4 keV and 2.0--3.5 keV, and 9.7--16.0 keV and 6.4--9.7 keV,
respectively.  The CCD is shown in Figure~\ref{fig-ccd}.  We observe
4U 1820-30 
in all three of the states associated with ``atoll'' sources (Hasinger
\& van der Klis 1989): the island state, lower banana branch, and
upper banana branch.  This confirms that the 176-day modulation is in fact
due to variation in the mass accretion rate.  Figure~\ref{fig-ccd}
shows that
we have adequately sampled
the long period and observed the source in all of its possible
states.  Closed symbols indicate data segments in which
kHz QPOs were discovered by Zhang et al. (1998), while filled symbols
indicate data without kHz QPOs.  

It is generally believed that in atoll sources the mass accretion rate
\mdot increases monotonically as the source moves from the island
state, through the lower banana, and into the upper banana branch
(Hasinger \& van der Klis 1989).  M\'endez et al. (1999) defined the
parameter $S_a$ to measure the position of an atoll source within its
CCD and thus parameterize its accretion rate.  We follow M\'endez et
al. (1999) and approximate the shape of the CCD track with a spline,
as shown in Figure~\ref{fig-ccd}.  $S_a$ is then defined as the
distance along this curve, as measured from the island state.  We
arbitrarily set $S_a = 1.0$ at (2.378,0.650) and $S_a = 2.0$ at
(2.671,0.597) in the CCD.  Each data point is assigned the value of $S_a$
of the point closest to it on the curve.  The error in $S_a$ is estimated by
simply projecting the error bars of the data points onto the curve.
Thus we may use $S_a$ in addition to the source luminosity to study
accretion rate-related changes in spectral parameters or kHz QPO
behavior.  The advantage in using $S_a$ instead of the hard color
(Kaaret et al. 1999) to
track the dependence of QPO frequency on accretion rate is that $S_a$
is more sensitive to changes of state along the nearly horizontal
banana branch, which is where the break in QPO frequency observed by
Zhang et al. (1998) occurs.

\subsection{Spectral Fitting}
\label{sec-fitting}

The spectral analysis was performed using XSPEC v10.0 (Arnaud 1996).
Spectra for 
each PCU and HEXTE cluster were reduced separately and combined within
XSPEC.  Due to systematic uncertainties in the response matrices, only
HEXTE data above 20 keV was used.  Several authors have written on
systematic effects in the PCA 
and their effect on spectral fitting (e.g. Sobczak et al. 1999; Barret
et al. 2000; Wilms et al. 1999; Tomsick et al. 1999); we performed
similar investigations and 
reached similar conclusions.  We extracted spectra for each PCU from
archival Crab observations spanning the year of our observations (1996
Oct 27, 1997 Mar 22, 1997 Jul 26, and 1997 Dec 15) and fit them with
an absorbed power law.  Initially we used response matrices generated
by {\em pcarmf} v3.5, but we got very poor fits with large
systematic deviations in the residuals, especially around 5 keV.  The
residuals were far better using the v2.2.1 response matrices for all
four Crab observations, and so
we used these for all our 4U 1820-30 fits.  The residuals still showed
large systematic effects below 2.5 keV and above 20 keV, and so we
restricted our fits to this energy range.  We also found, as have
other authors, that the residuals in PCUs 2 and 3 are considerably
worse than those in PCUs 0, 1, and 4.  For this reason we use only
PCUs 0, 1, and 4 in our fits.  PCU 4 was on during all observations
except 1997 May 28.  The
relative normalizations of the three PCUs and two HEXTE clusters were left as
free parameters, since there still exist large uncertainties in the
relative flux normalizations.  In our Crab fits, all three
PCUs find power law normalizations 20\%--30\% higher than the accepted
value of $\sim 10$ photons cm$^{-2}$ s$^{-1}$ keV$^{-1}$ at 1 keV
(e.g. Jung 1989).  Tomsick et al. (1999) have reported similar results.
All spectral model normalizations reported
here are obtained from PCU 0; in
Tables~\ref{tab-fits1}--~\ref{tab-fits3} (and also
Figure~\ref{fig-gauss}) we report the flux 
values found directly by PCU 0, while when comparing luminosities
with previous measurements (Figure~\ref{fig-comp}) we reduce the PCA values
by 15\% (see 
Section~\ref{sec-disc}).  
A systematic error of 1\% was added to
all PCA channels using {\em grppha}.  In Figure~\ref{fig-twospec} we
show two 
representative PCA and HEXTE spectra of 4U 1820-30, one from the low
island state and one from the high banana state.  The hardening of the
spectrum at low luminosity is clear.  In the island state
the source is detected by HEXTE only up to 50 keV, confirming the lack
of a detection above this by {\em BeppoSAX} (Piraino et al. 1999) and
BATSE (Bloser et al. 1996).  

For our spectral analysis data segments were grouped according to
their position in the CCD as follows:  Spectra were extracted for each
of the 90 data segments.  The four island state segments were combined
into one spectrum, and the five uppermost banana state segments were
each analyzed as separate spectra.  In between, data segments were
grouped into 26 bins of width 0.025 in $S_a$ and combined together,
with an 
average value of $S_a$ and the total exposure time computed for each bin.
Thus we had a total of 32 spectra to fit (see
Tables~\ref{tab-fits1}--~\ref{tab-fits3}).

All spectra were initially fit with two
models: the CPL 
model described in Section~\ref{sec-prev}, and the
recently-developed Comptonization model of Titarchuk (1994), called
the {\em CompTT} model in XSPEC.  This model improves on the CompST model by
including relativistic effects and extending the theory to work in
both the optically thin and thick cases.  In addition to the
Comptonizing electron temperature $kT_e$ and optical depth $\tau$, an
additional parameter is the temperature of the cool seed photons
$kT_W$, assumed to follow a Wien law.  We consider only a spherical
geometry for comparison with previous results.  In nearly all cases
the model fits were improved by the 
addition of a blackbody component.  We also included
a gaussian line component at $\sim 6.7$ keV in all spectra and a
smeared absorption edge in the island state, as described in more
detail in Section~\ref{sub-spec}.

Some XRB have recently been fit
with a multi-color disk blackbody (DBB; Mitsuda et al. 1984) instead
of a simple 
single-temperature BB to account for the soft excess (Guainazzi
et al. 1999; Barret et al. 1999).  We also fit all our spectra with a
model consisting of a CPL plus this disk blackbody  (CPL + DBB
model).  The DBB model parameters are the color temperature of the
inner disk $kT_{in}$ and the projected inner disk radius
$R_{in}\sqrt{\cos \theta}$, where $\theta$ is the inclination of the
system.  It is expected, however, that the true spectrum emitted by
the accretion disk will be that of a ``diluted blackbody'' due to the
effects 
of electron scattering, and the DBB parameters must then be modified
by a spectral hardening factor $f$ (Ebisawa et al. 1994).  The
effective temperature and inner radius are $kT_{eff} = kT_{in}/f$ and
$R_{eff} = f^2R_{in}$.  Shimura \& Takahara (1995) have found that $f
= 1.7$ is a good approximation for accretion parameters appropriate
for an XRB.  We attempted also to fit a CompTT + DBB model, but the
data were unable to constrain $kT_e$ and $\tau$ for this combination.

The insensitivity of the PCA below 2 keV did not allow us to
determine the hydrogen column density $N_H$ from our spectral fits.
It was therefore necessary to freeze $N_H$ at previously-determined
values for all spectra.  Optical reddening for NGC 6624 has been
measured by Liller \& Carney (1978) and Peterson (1993) to be $E_{B-V}
\sim 0.27$, which, using the canonical relationship $A_V = 3 E_{B-V}$,
yields an optical extinction $A_V = 0.81$.  Predehl \& Schmitt (1995)
find that $N_H = 1.79 \times 10^{21} A_V$ cm$^{-2}$, and this gives a
value of $N_H \sim 1.4 \times 10^{21}$  
cm$^{-2}$ for NGC 6624.  Previous X-ray observations of 4U 1820-30
with good low energy coverage 
({\em Einstein}, {\em BeppoSAX}; Smale et al. (1994) do not report a
value from {\em ASCA}) gave $N_H \sim 2.9 \times 10^{21}$ cm$^{-2}$,
and {\em EXOSAT} found $N_H \sim 4 \times 10^{21}$ cm$^{-2}$ (see
Table~\ref{tab-prev}).  Predehl \& Schmitt (1995) list values of
2.1--3.6 $\times 10^{21}$ cm$^{-2}$ for 4U 1820-30 from {\em ROSAT}
data.  These values are all in good agreement, and we therefore took a
rough average and set $N_H = 3.0 \times 10^{21}$ 
cm$^{-2}$ and kept it frozen for all our spectral fits.

\section{Results}
\label{sec-res}

\subsection{Saturation of QPO Frequency}
\label{sub-qpo}

The frequencies of the kHz QPOs reported by Zhang et al. (1998) are
plotted as a function of $S_a$, as derived here, in
Figure~\ref{fig-sa}.  Open and 
filled plotting symbols indicate the lower and higher frequency QPOs,
respectively, as defined by Zhang et al. (1998).  There is a clear
correlation of QPO frequency with $S_a$ below 
$S_a \sim 1.3$, and then the frequency saturates even though $S_a$
continues to rise.  This is an expected signature of the inner disk
reaching the marginally stable orbit (Miller, Lamb, \& Psaltis
1998). The QPOs 
disappear completely at $S_a \sim 1.45$.  These two ``special'' values
of $S_a$ are indicated in Figure~\ref{fig-ccd}.  If $S_a$, the position of
the source in the CCD, may indeed be taken to be an indicator of
\mdot, then Figure~\ref{fig-sa} provides strong evidence for the
detection of the marginally stable orbit in 4U 1820-30.  

We have also investigated the possibility that the shape of the CCD
varies with the 685-second orbital period.  The CCD was reproduced
with bins of 128 
seconds, and each point was grouped into one of five phase bins.  No
significant change in the track described by the CCD was found, as the
scatter in the points lay well within the error bars.  In addition, we
have confirmed that the shape of the frequency-$S_a$ plot in
Figure~\ref{fig-sa} does not become smeared out when smaller time bins
(and their associated larger error bars) are used.

\subsection{Evolution of Spectral Parameters}
\label{sub-spec}

The 32 combined PCA/HEXTE spectra were first fit with the CPL + BB
model with a gaussian line whose parameters were free to vary.  
Figure~\ref{fig-gauss} shows the gaussian centroid
energy $E_l$ and width ($\sigma$) as a function of the total luminosity
in the 2--50 keV band spanned by both instruments (see also
Table~\ref{tab-fits1}).  In all but the low
island state observation the gaussian parameters are consistent with
constant values of $E_l \sim 6.8$ keV and $\sigma \sim 0.8$ keV.
These are reasonable values for an Fe {\sc xxv} or {\sc xxvi} $K\alpha$ line
reflected from the inner regions of a hot disk (White et
al. 1986; Hirano et al. 1987).  The island state line energy is 6.53
keV, suggestive of reflection from less highly ionized gas, but the
difference is only $\sim 2\sigma$ and it is difficult to see how the
ionization could change so much for only a $\sim 14$\% change in
luminosity.  
We therefore froze $E_l$ at 6.8 keV and $\sigma$ at 0.8 keV 
for our fits with the CompTT + BB
model, since we found that the gaussian parameters were less well
defined with this choice of continuum.  The gaussian parameters were
left free for the CPL + DBB fits, and no difference was found between
the island and banana state line energies (Table~\ref{tab-fits3}).

Since a broad iron line suggests reflection from a disk, we
investigated the presence of other reflection components in our
spectra.  We first attempted to fit a smeared absorption edge (Ebisawa
et al. 1994; {\em smedge}
model in XSPEC) between 7 and 8 keV.  Only in the island state
spectrum could this component be fit with reasonable values.  Although
we would expect reflection to increase as the accretion rate and disk
area increase, this model requires a power law continuum to fit
against, and in all the rest of our observations the cutoff energy is
too low, resulting in too ``curvy'' a spectral shape for reliable
fitting of the edge with our data.  The edge energies in the island state
are $8.90^{+0.42 }_{-0.38}$ keV for the CPL + BB model and
$7.73^{+1.34}_{-0.36}$ keV for the CompTT + BB model, indicative again
of ionized material.  The smearing
width was frozen at 10 keV.  These edges are included in the island
state fits shown in Tables~\ref{tab-fits1} and \ref{tab-fits2}. The
smeared edge could not be fit with 
the CPL + DBB model.  We attempted next to fit the complete XSPEC reflection
model {\em pexrav} (Magdziarz \& Zdziarski 1995) to the island state
spectrum.  This model is the combination of a cut-off power law with a
Compton reflection component.  The derived reflection scaling factor
was consistent with zero for $\cos\theta = 0.45$.  We conclude that,
other than the broad iron line, we have only marginal evidence for
reflection from the disk, and only in the low state.

The best-fit spectral parameters for the CPL + BB model are given in
Table~\ref{tab-fits1}, the parameters for the CompTT + BB model
are given in Table~\ref{tab-fits2}, and the parameters for the CPL +
DBB model are given in Table~\ref{tab-fits3}.  In all three Tables we give
the average value of $S_a$ for the included data, and in
Table~\ref{tab-fits1} we list the total 
integration time for the PCA and 
HEXTE.  The CPL + BB spectral parameters are as in
Table~\ref{tab-prev}, with the addition of the effective radius of the
BB emitting surface $R_{BB}$ and the equivalent width of the
gaussian line.  The luminosity and flux ratio $L_{BB}/L_H$ are given
for the 2--50 keV band covered by the PCA and HEXTE.  (This is the
raw luminosity derived directly from the fit.)  All of the fits
are acceptable, and in many cases the $\chi^2_{\nu}$ is so low as to
indicate that the systematic errors have been overestimated.  In
Table~\ref{tab-fits2} we give the best-fit temperature of the seed
photons $kT_W$ and the Comptonizing $y$-parameter, defined as $y =
4kT_e\tau^2/m_ec^2$.  We also follow in 't Zand et al. (1999) and
derive the effective Wien radius $R_W$ of the seed photons:
the total flux in the CompTT component is corrected for energy gained by
Comptonization by dividing by $1 + y$, and this is set equal to the Wien flux
from the surface of a sphere.  The expression used is 
$R_W = 3 \times 10^4 D \sqrt{F_{CompTT}/(1 + y)}/(kT_W)^2$ km, where
$D$ is the distance in kpc, $F_{CompTT}$ is in erg cm$^{-2}$ s$^{-1}$,
and $kT_W$ is in keV.  The derived values are listed in
Table~\ref{tab-fits2}. 
The quality of the CompTT + BB fits is
practically identical to that of the CPL + BB fits; based on the
values of $\chi^2_{\nu}$ there is no reason to favor one model over
the other.  We note however that several of the CompTT + BB fits have
very large errors on the BB parameters.  In Figure~\ref{fig-fit} we
show two examples of spectra from PCUs 0, 1, and 4 and HEXTE clusters A
and B fit with the CompTT + BB model.  These
are the same two spectra shown in Figure~\ref{fig-twospec}; now we
show the raw counts with the folded model, the residuals, and the
unfolded spectra with each component plotted separately.  We note that
the ``wave'' shape seen in the residuals of PCA spectra around 6 keV is
still visible.  In Table~\ref{tab-fits3} we include the fitted disk
parameters $kT_{in}$ and $R_{in}\sqrt{\cos\theta}$ with the standard
CPL and gaussian parameters.  The fits are again all formally
acceptable.  We note, however, that in the island state the model is
unable to reproduce the measured counts above 40 keV.

In Figure~\ref{fig-params} we show the spectral
parameters of the CPL + BB and CompTT + BB models as a function of
$S_a$, and thus presumably of \mdot.   In both models it is clear that
the parameters of the hard 
spectral component are correlated with the accretion rate, especially
as the accretion rate become low, in the sense that the spectrum
becomes harder as \mdot decreases.  The cutoff energy $E_c$ in the CPL model
and $kT_e$ in the CompTT model both rise as the
accretion rate drops.  Note that when fitting a CPL model it
is not
$\alpha$ but $E_c$ that indicates the hardness of the spectrum, since
$\alpha$ here is really an indication of the optical depth.
This change is not gradual throughout the CCD,
however; only as 
the source moves from the lower banana to the island state do the
parameters change appreciably.  The optical depth $\tau$ of the CompTT
model drops at lower \mdot, and the power law index $\alpha$
(inversely related to $\tau$ in the CPL approximation) rises.
This general behavior is consistent with that seen in other sources
(e.g., Barret et al. 2000; Christian \& Swank 1997).  

The most intriguing feature of Figure~\ref{fig-params} is that the
parameters of the CompTT + BB model show a distinct change at $S_a
\sim 1.45$.  The ratio $L_{BB}/L_H$ is inversely correlated with $S_a$
below this, but becomes constant above.  The blackbody temperature
$kT_{BB}$ drops from $\sim 2.3$ keV to $\sim 1.6$ keV, and the
blackbody radius $R_{BB}$ begins to increase from $\sim 1$ km to $\sim$ 2
km.  The $\tau$ versus $S_a$ curve also seems to roll over and become
nearly constant at this value of $S_a$.  In fact, $\tau$ and
$L_{BB}/L_H$ appear strongly anti-correlated below $S_a = 1.45$.  No
such abrupt change is seen 
in the parameters $L_{BB}/L_H$, $kT_{BB}$, $R_{BB}$, or $\alpha$ of
the CPL + BB model.  The cutoff parameters $E_c$ in the CPL model and
$kT_e$ in the CompTT model do not seem to change at $S_a = 1.45$;
however, both appear to change from strong anti-correlations with
$S_a$ to near-constant values at $S_a \sim 1.3$.  These ``special''
values of $S_a = 1.3$ and $S_a = 1.45$ are especially interesting
because, as shown in
Section~\ref{sub-qpo} above, these are the same values at which the
kHz QPOs saturate in frequency and disappear completely,
respectively.  
Thus the accretion flow must be changing at the values
of \mdot corresponding to these values of $S_a$ in a way that affects
both the energy spectrum and the frequencies of the kHz QPOs.  The
fact that this correlation is seen only with the CompTT + BB model
suggests that this model is the more accurate description of the
spectrum. 
In Figure~\ref{fig-wien} we show the seed photon temperature $kT_W$ and the
Wien radius $R_W$ described above as a function of $S_a$.  Although
the error bars are large, there is an indication that the temperature
and radius are anticorrelated below $S_a = 1.3$ and constant above.
This is precisely the behavior expected if the seed photons are coming
from the inner part of the accretion disk, and the disk is contracting
until it reaches the last stable orbit at $S_a = 1.3$ as indicated by
the kHz QPOs.  This trend gives us confidence in the fitted values of
$kT_W$, which must still be treated with some caution because the peak
of the Wien spectrum at $3kT_W$ lies below the PCA energy range.  The
large values of the Wien radius also suggest that 
the seed photons originate in the disk. 

The parameters for the CPL + DBB model are shown in
Figure~\ref{fig-paramsdbb}.  The cut-off power law components behave
similarly to those in the CPL + BB model, though the values of
$\alpha$ and $E_c$ are both considerably lower.  This is because the
DBB model prefers lower temperatures than the simple BB, and so the
curvature in the spectrum at higher energies must now be described by
a lower cutoff value.  The behavior of the DBB component is of great
interest: $kT_{in}$ rises and $R_{in}$ falls with increasing accretion
rate until $S_a \sim 1.3$, then both remain nearly constant (though there
is some indication that $kT_{in}$ continues to rise slowly).  The fraction
of the total luminosity in the DBB component increases with $S_a$ in
the same manner.  Thus it appears that the accretion disk inner radius
grows smaller and hotter as the accretion rate increases until the
last stable orbit is reached at $S_a = 1.3$.  This behavior of the DBB
component is consistent 
with that expected from the behavior of the kHz QPOs.  Thus this model
is also compelling as an accurate physical description of the system.

\section{Discussion}
\label{sec-disc}

\subsection{Comparison with Previous Spectral Measurements}

First we compare our spectral results for 4U 1820-30 to those
previously reported using the CPL + BB and CompTT + BB models.
In Figure~\ref{fig-comp} we show our {\em RXTE} spectral parameters
together 
with the previously-reported parameters described in
Section~\ref{sec-prevspec} and Table~\ref{tab-prev}.  Here we plot the
spectral parameters as a function of the total 2--50 keV luminosity,
assuming a distance of 6.4 kpc.
The luminosities of the previous measurements have
been translated into the 2--50 keV band based on the reported
parameters and fluxes.  The integrated 2--50 keV flux in the archival
Crab spectra (see Section~\ref{sec-fitting}) is $\sim 15$\% higher
than what is expected; therefore,
the {\em RXTE} luminosities in Figure~\ref{fig-comp} have been divided
by a factor of 1.15 for proper comparison to previous instruments.
It is immediately obvious that there is considerable 
disagreement between different instruments.  In general, the {\em
RXTE} spectral parameters agree most closely with those found by {\em
EXOSAT} and {\em Ginga}, and differ the most from those found by {\em
BeppoSAX}.  More specifically, the {\em RXTE} BB component parameters agree
fairly well with those found by {\em EXOSAT} and {\em Ginga} for the
CPL + BB model and with those found by {\em EXOSAT} for the CompTT +
BB model.  In the CompTT + BB model, for example, {\em EXOSAT} also
observed $kT_{BB}$ to 
decrease from $\gtrsim 2$ keV to $\sim 1.5$ keV between the low and high
states, while $R_{BB}$ increased from $\sim 1.7$ km to $\sim 3.7$ km.
The {\em EXOSAT} CPL parameters also agree well with our values in the
low and high states, as do the {\em Einstein} values (Christian \&
Swank (1997) do not report BB values for the CPL model with {\em
Einstein}).  The {\em BeppoSAX} BB parameters and CPL spectral index
$\alpha$ differ considerably from ours. All of the experiments that
used a CompST model found 
Comptonization parameters that agree well with our CompTT parameters at
high luminosity.  In the low state, however, {\em EXOSAT} did not
observe the increase in $kT_e$ and decrease in $\tau$
that we see.  It is strange that the {\em EXOSAT} low state observation
could only place a lower limit of $\sim 30$ keV on $E_c$ for the CPL
model, yet could still constrain such a low $kT_e \sim 3.5$ keV for
the CompST model.  It is apparent that our island state observation is
the first observation of 4U 1820-30 in the low state since {\em
EXOSAT}, and it is puzzling that the Comptonization parameters
differ so much.  Further observations in the island state are needed, but
the low duty cycle in this state seen in Figure~\ref{fig-asm}
makes this difficult.

It is apparent from Table~\ref{tab-prev} and Figure~\ref{fig-comp}
that those experiments without sensitivity below 1 keV, i.e. {\em
EXOSAT}, {\em Ginga}, and {\em RXTE}, find consistently higher
$kT_{BB}$ and lower $R_{BB}$ than do those instruments with
sensitivity below 1 keV, i.e. {\em ASCA}, {\em Einstein}, and
{\em BeppoSAX}.  The differences are important to the question of the
origin of the soft component of the spectrum.  The instruments with
low energy sensitivity find $R_{BB} \gtrsim 10$ km, consistent with the
radius of a neutron star.  The other instruments, including {\em
RXTE}, find $R_{BB} \lesssim 4$ km, implying that the blackbody component can
come from only a limited region of the neutron star surface, such as
an optically thick boundary layer.  (An alternate explanation for
apparently small blackbody radii for XRBs given by
Rutledge et al. (1999), that 
a blackbody is not an appropriate model and fits with hydrogen
atmosphere models give radii near 10 km, is not applicable to
4U 1820-30 since the accretion rate is too high.)  We have attempted
to fit our {\em 
RXTE} spectra with a BB component with $kT_{BB} < 1$ keV, forcing the
fit to find a new local minimum.  
For a CompTT + BB model of this type, $kT_{BB}$ remained nearly
constant at $\sim 0.6$ keV while $R_{BB}$ stayed at $\sim 24$ km,
too large for a NS surface.  The seed photon energy $kT_W$
decreased 
from 1.8 keV to 1 keV as $S_a$ increased from 1.0 to 1.4, which also
seems unphysical, and neither $kT_e$ nor $\tau$ changed their behavior
at the interesting values of $S_a$.  
The new model was unable to reproduce the
observed counts at energies above 30--40 keV in the island state.
This, together with the 
clear change in the behavior of our BB parameters at an
``interesting'' value of $S_a$ (or $L$) seen in
Figure~\ref{fig-params}, leads us to conclude that the BB component
parameters listed in Tables~\ref{tab-fits1} and~\ref{tab-fits2} are
the best fits to our data for the CPL + BB and CompTT + BB models.
We note, however, that properly measuring low energy components
depends on having the correct $N_H$, which we cannot measure with the
PCA, and so it is entirely possible that additional 
components exist in the true spectrum.  This notion is perhaps
supported by the fact that the seed photon temperature
found in the CompTT + BB model is in good agreement with the BB
temperature found by {\em BeppoSAX}.

\subsection{Physical Interpretation of the CompTT + BB and CPL + DBB
Models} 

Our {\em RXTE} data on 4U 1820-30 may be described equally well by two
different physically-motivated models (CompTT + BB and CPL + DBB), both
of which have parameters that appear to 
``know'' about the saturation of kHz QPO frequency at the accretion
rate corresponding to $S_a = 1.3$.  The physical interpretation of
these two models leads to two different pictures of the origin of the
spectral components in XRBs.  Neither picture, however,
is completely consistent with theoretical and observational
constraints.

The question of the origin of the soft and hard spectral components
in XRBs has been addressed by many authors (e.g., White, Stella, \&
Parmar 1988; Guainazzi et al. 1998; Olive et al. 1999, Barret et
al. 1999, 2000).  Early work supposed that the accretion disk
intersects the NS surface in an optically thick boundary layer which
is the source of the blackbody component observed in XRBs.
A potentially useful diagnostic is the ratio of
flux or luminosity in the BB component to the flux in the hard
component ($L_{BB}/L_H$).  Sunyaev \& Shakura (1986) showed that, if
relativistic effects are taken into account, the luminosity of the
boundary layer should be at least equal to that of the disk, and can
be more than twice it if the disk extends all the way to the
marginally stable orbit.  We find $L_{BB}/L_H < 0.5$ at
all times for 
the CompTT + BB and CPL + DBB models, and in the CompTT + BB model
at high accretion rates it is as low as 0.1.  
Many authors (White, Stella, \&
Parmar 1988; Barret et al. 1999, 2000; Table~\ref{tab-prev}) have
found similar results 
for this and other sources.  Guainazzi et al. (1998) and Barret et
al. (1999) suggest that this 
implies the weak BB component is produced by an optically thick accretion
disk, while the
strong CompTT component originates in an optically {\em thin} boundary layer
through a process such as the gap accretion model of Klu\'zniak \&
Wilson (1991).  Barret et al. (2000), however, argue that the
similarities between NS and BH hard X-ray emission require that both
be produced by the same mechanism, which rules out boundary layer
processes.  In addition, Church et al. (1998) have found that the
evolution of spectra during X-ray dips in sources such as 4U 1916-05
are consistent with a compact blackbody emitter being occulted while
an extended hard X-ray emitter is only partially covered.  From timing
of dip ingresses and egresses they estimate a diameter of $\sim 4
\times 10^{9}$ cm for the hard X-ray emitting region.  Thus one
possibility is that the BB component is produced in an optically
thick boundary layer covering a fraction of the neutron star surface,
while the CPL or CompTT component comes from an extended hot corona.
Some mechanism must then reduce the luminosity of the boundary layer
relative to that of the disk and corona.  White, Stella, \& Parmar
(1988) suggest that the NS may be spinning close to equilibrium with
the inner edge of the accretion disk, which would lessen the energy
dissipated in the boundary layer, but the NS spin periods measured
with {\em RXTE} (typically $\sim 300$ Hz) are well below this.
Another possibility discussed by Barret et al. (2000) is that an
optically thick boundary layer does not exist.  Instead, within the
standard optically thick accretion disk there exists a
hot, optically thin phase of the accretion flow such as the
advection-dominated solutions of Narayan \& Yi (1995).  Such a picture
has been applied successfully to BH systems (Esin et al. 1998).  Barret et
al. (2000) suggest that the boundary between such a hot flow and the
NS may well be optically thin and thus neither contribute to the BB
component (now coming entirely from the disk) nor quench the hot
advective flow and cause it to collapse to a thick disk (Yi et
al. 1996).  The NS 
surface must still be a source of cool reprocessed photons, however, and it
is not clear what effect this must have.

We can interpret our CompTT + BB model fits for 4U 1820-30 in terms of
the former of 
these two possibilities, in which the 2.3 keV BB flux comes from an
optically think boundary layer and the CompTT component from an
extended corona.
We have noted that in Figure~\ref{fig-params} the ratio
$L_{BB}/L_H$ decreases as the optical depth of the Comptonizing cloud
increases, and both become nearly constant at the same accretion
rate.  In Figure~\ref{fig-tauratio} we show $L_{BB}/L_H$ as a function
of $\tau$; the two are obviously strongly anticorrelated.  Thus
perhaps the BB emission from the boundary layer is simply 
becoming more obscured as the optical depth of the corona increases,
and the boundary layer emission is being redistributed to higher
energies by the Comptonizing cloud.  At the accretion rate
corresponding to the break, the $\sim 2.3$ keV boundary layer could be
completely obscured, and a cooler, $\sim 1.7$ keV disk component
would begin to dominate the BB part of the spectrum.  This component is
emitted from radii outside the inner hot cloud, leading to the upward
trend in $R_{BB}$ observed at the same $S_a$.  In this scenario a
multicolor disk blackbody model would be preferable for the higher
luminosity spectra, but as noted in Section~\ref{sec-fitting} fits of
a DBB together with the CompTT model were indeterminate.

Such a scenario ties in with the accretion rate-dependence of the kHz
QPO frequencies.  The origin of kHz QPOs is often explained by means
of a ``beat-frequency'' model.  The magnetospheric beat-frequency
model (Lamb 1991), which was developed to explain horizontal branch
oscillations (HBOs) in Z sources, has been suggested by Strohmayer et
al. (1996) as the 
source of kHz QPO pairs as well.   Miller, Lamb, \& Psaltis (1998)
point out that it is not clear how the X-ray modulation at the orbital
frequency of the disk is produced in this model, and that it cannot
explain both the HBOs and the kHz QPOs sometimes seen together in Z
sources.  Instead, Miller, Lamb, \& Psaltis (1998) propose the
sonic-point beat-frequency model, in which the disk is 
truncated by radiation drag from the NS which causes clumps of gas to
lose energy and fall rapidly inward on spiral trajectories.  This
spiral pattern crashes onto the star to create X-ray hot spots which
rotate at the frequency of the disk inner edge, generating the high
frequency QPO.  This flow is optically thin, which permits the
radiation to exert its drag on the disk in the first place.  At the
same time, some gas is channeled by the NS magnetic field onto hot
spots which rotate at the spin frequency of the star.  These hot spots
modulate the radiation drag, and thus the brightness of the first hot
spots, at the frequency of the beat between the disk orbital and stellar
spin frequencies, producing the second QPO.  Miller, Lamb, \& Psaltis
(1998) point out several predictions of this model as the inner edge
of the disk, or ``sonic point,'' approaches the marginally stable
orbit.  First, the frequencies of the QPOs should increase together up
to a certain accretion rate, then remain constant.  This is clearly
seen in Figure~\ref{fig-sa}.  Second, after the marginally stable
orbit is reached, the amplitudes of the QPOs should decrease.  This
is because, as \mdot increases further, the optical depth between the
sonic point and the NS surface will increase, lessening the radiation
drag which produces the modulation in the first place.  Zhang et
al. (1998) found that the fractional rms amplitudes of the QPOs
steadily decreased as the count rate increased until they disappeared
altogether.  

Thus we can envision the following scenario:  In the island state the
inner edge of the disk lies at a  radius greater than $6GM/c^2$,
within which the 
sonic-point model is operating.  The boundary layer is optically
thick, producing blackbody radiation from a small equatorial region of
the NS surface.  The NS is surrounded by a hot cloud in the
form of an accretion disk corona, since the optically thick boundary
layer prevents an advection-dominated solution from existing in this
case.  The Wien temperature and radius also indicate that the corona
is supported by the disk.  The electron scattering optical
depth of the corona is low, and we may see the boundary layer
emission.  As \mdot increases the inner edge of 
the disk moves inward and the QPO frequencies increase.  The optical
depth of the corona increases and, as the boundary layer and disk emission
increase, the corona is cooled and $kT_e$ drops.  The increased
emission from the boundary layer does not make it out to the observer,
as it goes into cooling the corona.  The derived Wien radius also
indicates that the disk inner radius is shrinking.  When the inner
edge of the disk 
reaches the marginally 
stable orbit, the QPO frequencies cease rising.  The disk has reached
its maximum emitting area for cool photons, and so $kT_e$ does not
fall any more.  Further
increasing the accretion rate causes the optical depth of both the corona
and the spiraling gas clumps between the disk edge and the boundary
layer to rise, reducing the radiation drag that ultimately generates
the QPOs.  The QPO amplitude begins to decrease until, at the point
when the boundary layer is completely obscured, they disappear
altogether.  At this point the cooler emission from the disk itself is
all that can be seen from the outside, resulting in a lower observed
$kT_{BB}$ and higher $R_{BB}$, as the boundary layer powers the
optically thick, nearly thermal-looking Comptonizing cloud.  The ratio
of BB to CompTT flux is quite small at this point.  

The main problem with this picture is that the Comptonizing corona is
supported by the accretion disk, as indicated by the Wien parameters
(Figure~\ref{fig-wien}), and yet its luminosity is always more than
twice that of the BB component.  If the boundary layer is being
obscured by the corona, transferring energy from the soft to the hard
component, then the seed temperature of the CompTT
model should start to reflect this.  Thus if the corona is supported
by the disk this picture is not
consistent with theoretical constraints on the ratio of the flux from
the boundary layer and disk (Sunyaev \& Shakura 1986).

Our CPL + DBB model for 4U 1820-30 may be interpreted in terms of a
hot advective 
flow lying within the standard thin accretion disk if the boundary
layer between this hot, optically thin flow and the NS may also be
optically thin (Barret et al. 2000).  The scenario is fairly
straight forward in this case.  In the island state, the inner disk
radius is large and the temperature is low.  Between the disk and the
NS the hot advective flow Comptonizes soft photons from the disk and
NS surface, producing the CPL component.  As the disk lies outside the
hot corona, it is clearly visible.
As \mdot increases, the
inner disk moves inward and its temperature increases, as shown in
Figure~\ref{fig-paramsdbb}.  As the disk area increases, the ratio of
DBB to CPL flux increases, as does the cooling of the hot flow,
causing the cutoff energy to drop.  The power law index falls as well,
indicating an increase in $\tau$.  
The QPOs may be produced by the same
sonic point model as before, and so their frequency increases.  When
the inner disk reaches the last stable orbit, the flux ratio,
temperature, radius, optical depth, and cutoff energy reach constant
values.  As \mdot continues to rise, $kT_{BB}$ does appear to
continue rising slowly as well.  

We may use the DBB parameters at the last
stable orbit to estimate the mass of the NS, as described by Ebisawa
et al. (1994) and Shimura \& Takahara (1995).  As discussed in
Section~\ref{sec-fitting}, $R_{in}$ must be corrected by a spectral
hardening factor; this effective radius is then set to the radius at
which the disk temperature becomes a maximum, 5/3 the inner disk
radius itself.  Then $3 R_S = (3/5) \times 0.6 \times R_{in}
f^2$, where $R_S$ is the Schwarzschild radius.  The factor of 0.6
accounts for the decrease in 
$R_{in}\sqrt{\cos \theta}$ due to relativistic effects.  Using $f =
1.7$ from Shimura \& Takahara (1995) and $R_{in} = 8$ km from
Figure~\ref{fig-paramsdbb} gives a NS mass of $M_{NS} =
0.94/\sqrt{\cos \theta}$ M$_{\odot}$.  The NS mass may also be
estimated by setting the maximum kHz QPO frequency of $\sim 1050$ Hz
equal to the Keplerian frequency at the last stable orbit.  This gives
$M_{NS} = 2.1$ M$_{\odot}$ for a non-rotating NS.  For a NS spin
frequency of 275 Hz (Zhang et al. 1998) the relativistic corrections
are small.  Equating these two masses requires $\theta = 78^{\circ}$,
which is nearly problematic due to the lack of dips from this source.
Should any DBB flux be lost due to scattering in the hot corona,
however, the true value of $R_{in}$ and thus $M_{NS}$ would appear
reduced. 

Appealing as this physical model is, it is inconsistent with the
observational work
of Church et al. (1998), who find that X-ray dips in XRBs may be modeled by
partial absorption of a large corona responsible for the hard X-ray
emission.  They estimate a diameter of $4 \times 10^{9}$ cm for this
corona, considerably larger than the $16f^2 = 46$ km $= 4.6 \times 10^{6}$ cm
allowed from 
Figure~\ref{fig-paramsdbb} if the hot advective flow is to lie inside
the accretion disk.  Tomsick, Lapshov, \& Kaaret (1998) analyze a dip
observed from the BH candidate 4U 1630-47 and find that the power law
component of the spectrum is more highly absorbed, in contrast to the
situation found by Church et al. (1998) for XRBs.  Thus a hot
advective Comptonizing flow lying within the accretion disk may be a
good model for BH systems, but seems to present a geometry problem for
XRBs.

\subsection{Lack of Extensive Hard X-ray Emission: NS vs BH Systems}

As noted in Section~\ref{sec-prevspec}, 4U 1820-30 does not belong to
the class of XRBs detected at $\sim 100$ keV.  Our {\em RXTE}
observations confirm this, as the source is not detected above 50 keV
even in the hardest state.  Extrapolating the island state spectrum
from Table~\ref{tab-fits2} into the BATSE energy range (20--100 keV)
gives a hard X-ray luminosity of $\sim 2.5 \times 10^{36}$ \ergs,
consistent with the upper limit of $\sim 2.8 \times
10^{36}$ \ergs 
found by BATSE (Bloser et al. 1996).  The predicted flux at 100 keV
in the island state is $\sim 2 \times 10^{-7}$ photons cm$^{-2}$ s$^{-1}$
keV$^{-1}$. From Figure~\ref{fig-asm} is is
clear that the source spends only a small fraction of its duty cycle
in the island state, making hard X-ray detection even less likely.
Barret, McClintock, \& Grindlay (1996; see also Barret et al. 2000)
have proposed that XRBs can only produce substantial hard X-ray
emission when their soft X-ray (1--20 keV) luminosity is below a
critical value of $\sim 1.5 \times 10^{37}$ \ergs, and that this may
be used to distinguish NS systems from BH systems.  Since the 1--20
keV luminosity of 4U 1820-30 never falls below $\sim 2 \times 10^{37}$
\ergs, even in the island state, our lack of detection of the source
above 50 keV is consistent with this picture.

\section{Conclusions}
\label{sec-conc}

We have studied the X-ray spectrum of 4U 1820-30 in all three states
displayed by atoll sources, including the first observation in the
island state since the {\em EXOSAT} observation described by Stella,
White, \& Priedhorsky (1987).  The fact that the 176-day flux
variation corresponds to normal motion in the CCD indicates that the
long period is indeed due to modulation in the accretion rate.
The spectra are well fit by the
CPL + BB, CompTT + BB, and CPL + DBB models, but the parameters of the
CompTT + BB and CPL + DBB
models show more robust correlations with each other and with the
frequencies of the kHz QPOs when plotted as a function of accretion
rate (as parameterized by $S_a$).  
In the CompTT + BB case, the relative contribution of the
blackbody component decreases as the optical depth of the Comptonizing
cloud increases until both reach constant values at the same \mdot at
which the QPOs disappear.  This can be understood in terms of simple
geometrical arguments and the sonic-point beat-frequency QPO model of
Miller, Lamb, \& Psaltis (1999) if the blackbody component emanates
from an optically thick boundary layer and the hard Comptonized
component from an extended cloud; however, the relative fluxes of the
two components violate theoretical constraints. 
In the CPL + DBB case, the behavior of the parameters for the inner accretion
disk agree with expectations from the QPOs, but the Comptonizing
corona must take the form of a hot advective flow within the standard
thin disk; observations of dipping XRBs suggest the corona is much
larger than this.  The questions raised by both models are central to
the issue of hard X-ray production in NS systems.  In either case,
the saturation of QPO frequency at
$S_a = 1.45$ provides strong additional evidence for the detection of
the marginally stable orbit in the accretion disk of 4U 1820-30.

\acknowledgments
We would like to thank Dimitrios Psaltis for early discussions of the
data.  This paper had made use of 
quick-look results provided by the ASM/{\em RXTE} team.  This work was
supported in part by NASA grant NAG5-7393.
PFB acknowledges support from
NASA GSRP grant NGT5-50020.

\clearpage

\figcaption[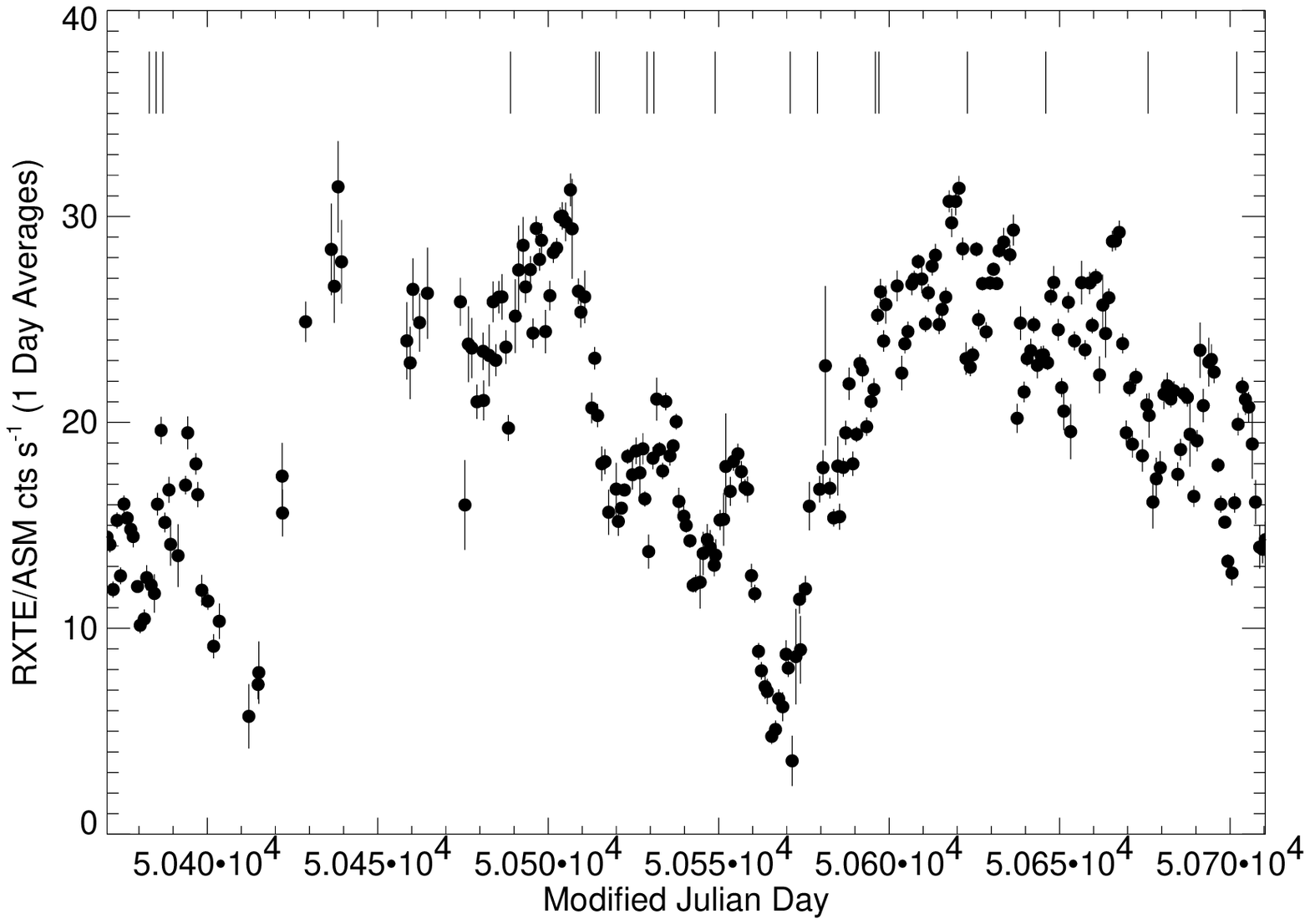]{{\em RXTE}/ASM light curve of 4U 1820-30
(2--12 keV) from 
1996 October -- 1997 October (MJD 50400 = 1996 November 12).  Each
point represents a 1-day average.  Pointed PCA/HEXTE
observation times are marked 
with vertical lines.  The $\sim 176$ day modulation is clear.
\label{fig-asm}}

\figcaption[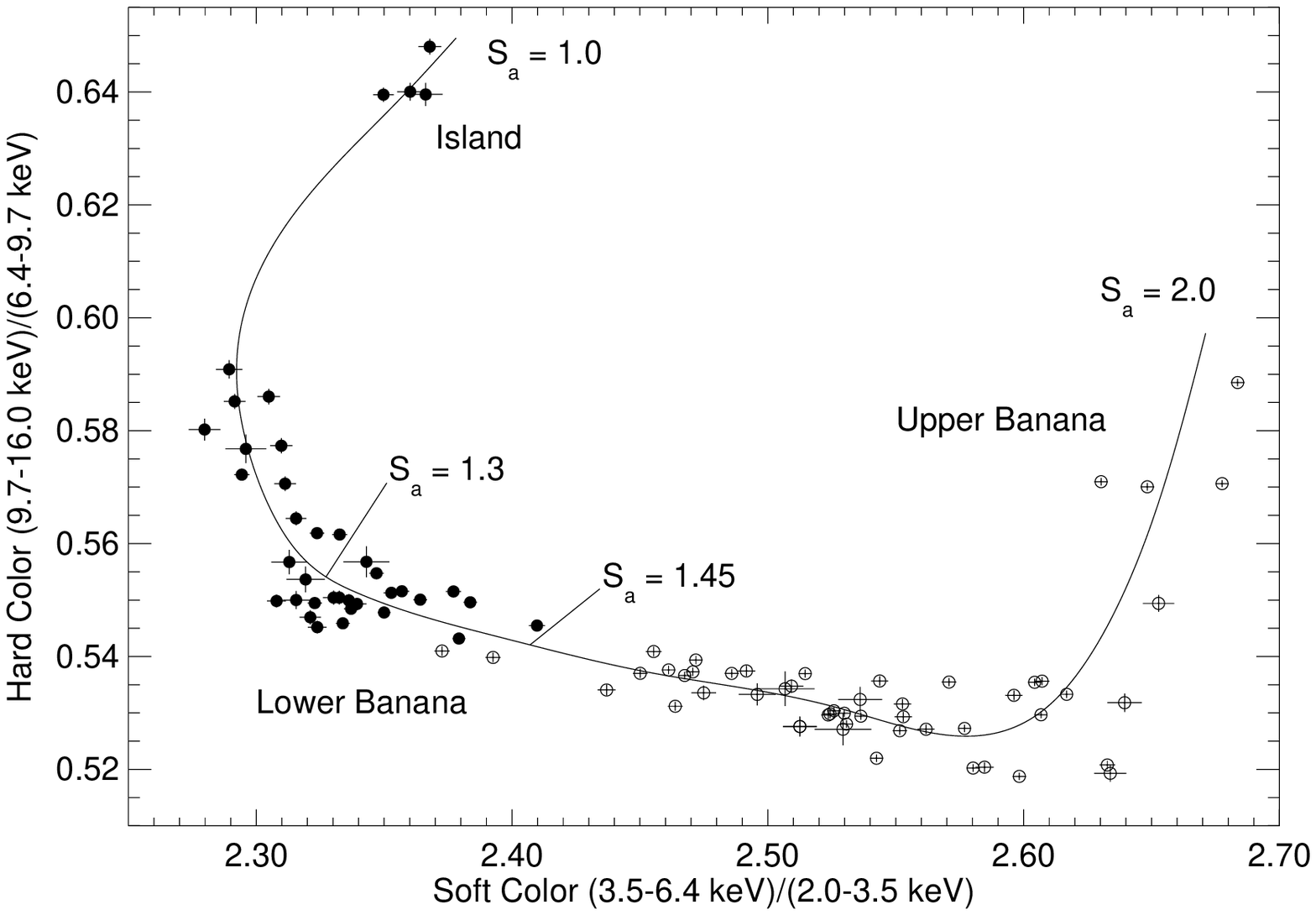]{Color-color diagram of 4U 1820-30.
Each point represents one 
of the 90 data segments from Zhang et al. (1998), which have typical lengths
of 3000 s.  The soft and hard colors are defined as the ratio of
background-subtracted count
rates in the bands 3.5--6.4 keV and 2.0--3.5 keV, and 9.7--16.0 keV
and 6.4--9.7 keV, respectively.  Filled symbols indicate data
that contain kHz QPOs from Zhang et al. (1998), open symbols data that
do not.  The position within the diagram is parameterized by the
variable $S_a$, the distance along the fitted line.  Important values of
$S_a$ are indicated in the Figure for clarity.
\label{fig-ccd}}

\figcaption[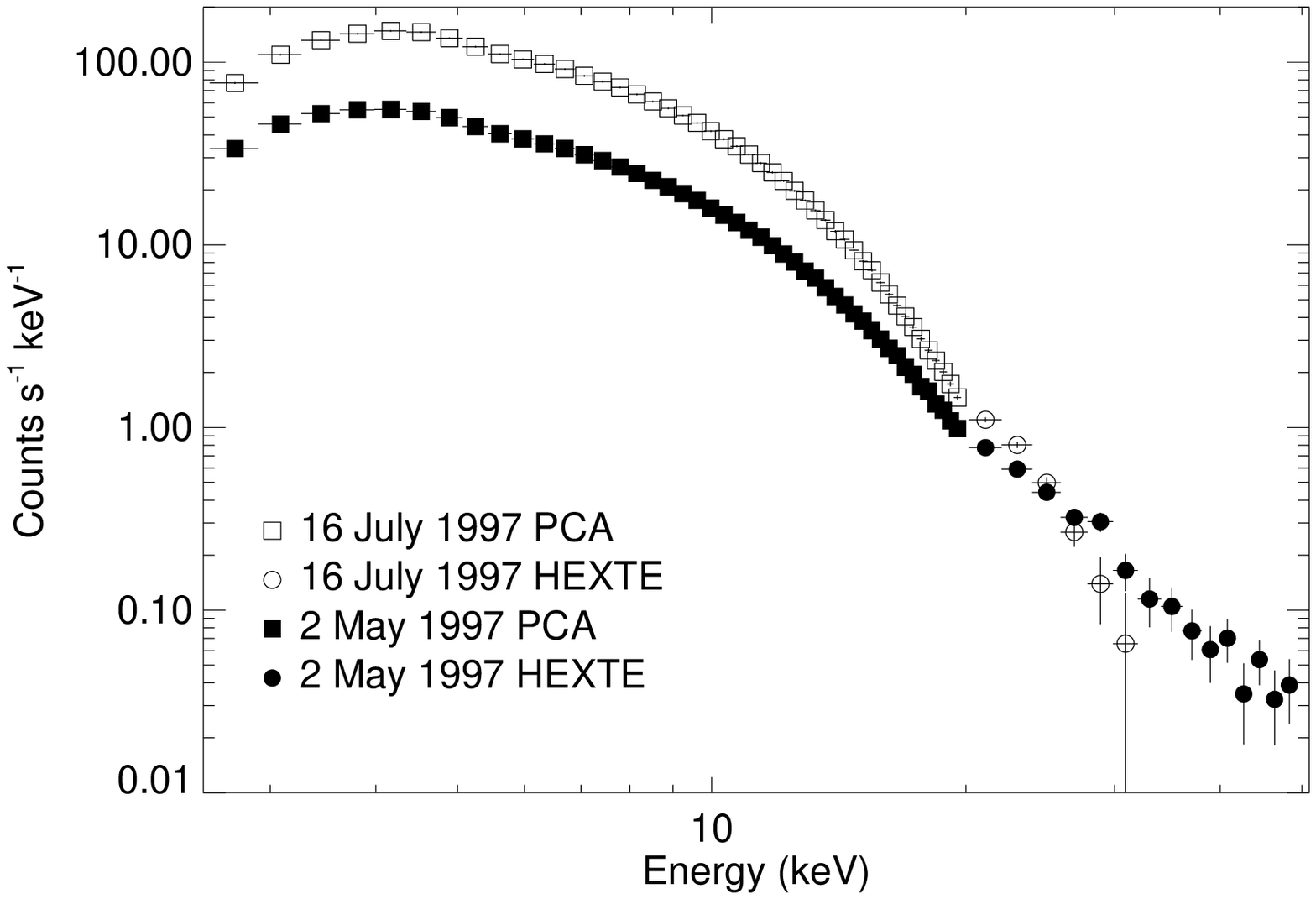]{Representative PCA (PCU 0) and HEXTE
(cluster A) spectra of 4U 1820-30.  On 1997 
July 16 the source was in a high state, in the upper banana portion of
the color-color diagram.  On 1997 May 2 the source was in the low
island state and is detected with HEXTE up to 50 keV.  The hardening
of the spectrum with decreasing luminosity is clear.
\label{fig-twospec}}

\figcaption[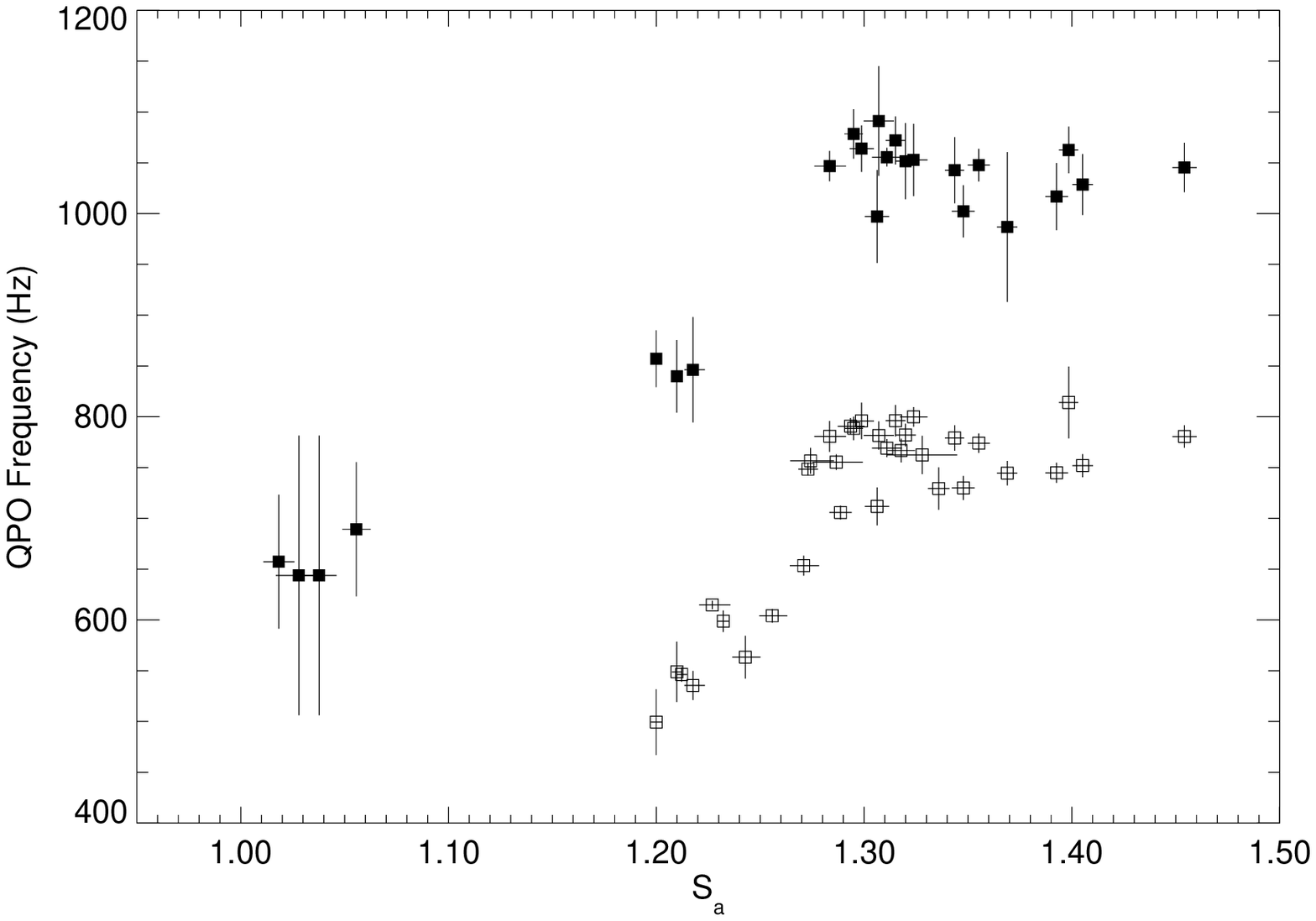]{Relation between kHz QPO frequency and
position within the 
color-color diagram, measured by the parameter $S_a$.  Filled and open
symbols represent the high and low frequency QPOs, respectively.
There is a clear saturation in QPO frequency even though $S_a$, and
thus presumably the accretion rate, continues to rise.
\label{fig-sa}}

\figcaption[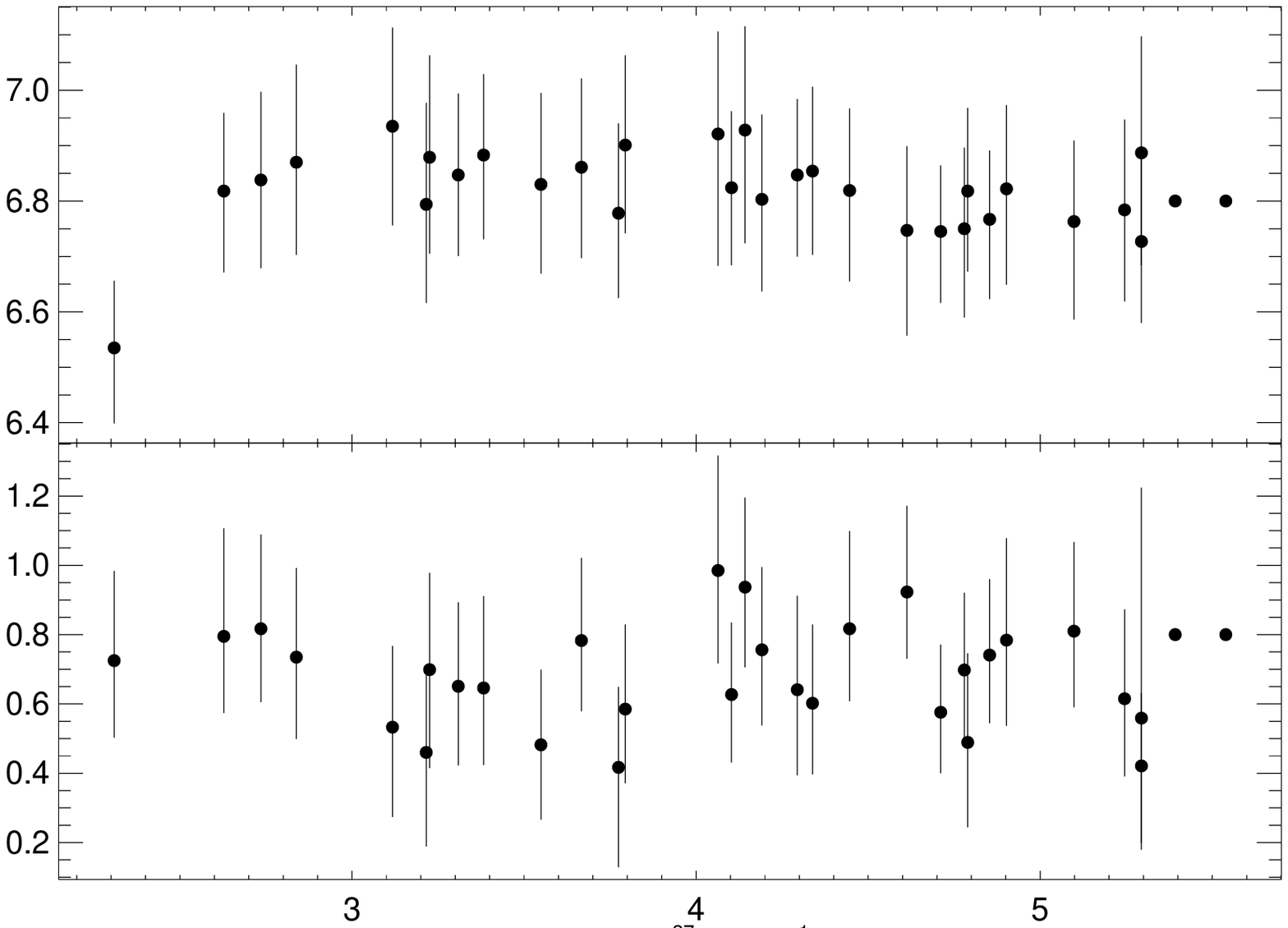]{Fitted centroid energy and width
($\sigma$) of the 
gaussian line as a function of total luminosity (2--50 keV) for the
CPL + BB model.  With the 
exception of the island state observation, the centroid energy remains
constant at $\sim 6.8$ keV and the width remains constant at $\sigma
\sim 0.8$ keV, indicative of reflection from ionized gas in the disk.
The line energy in the island state, $\sim 6.55$ keV,  suggests
less ionized gas, but the difference in only $\sim 2\sigma$.  A
smeared absorption edge at 8.9 keV is included 
in the island state fit.
\label{fig-gauss}}

\figcaption[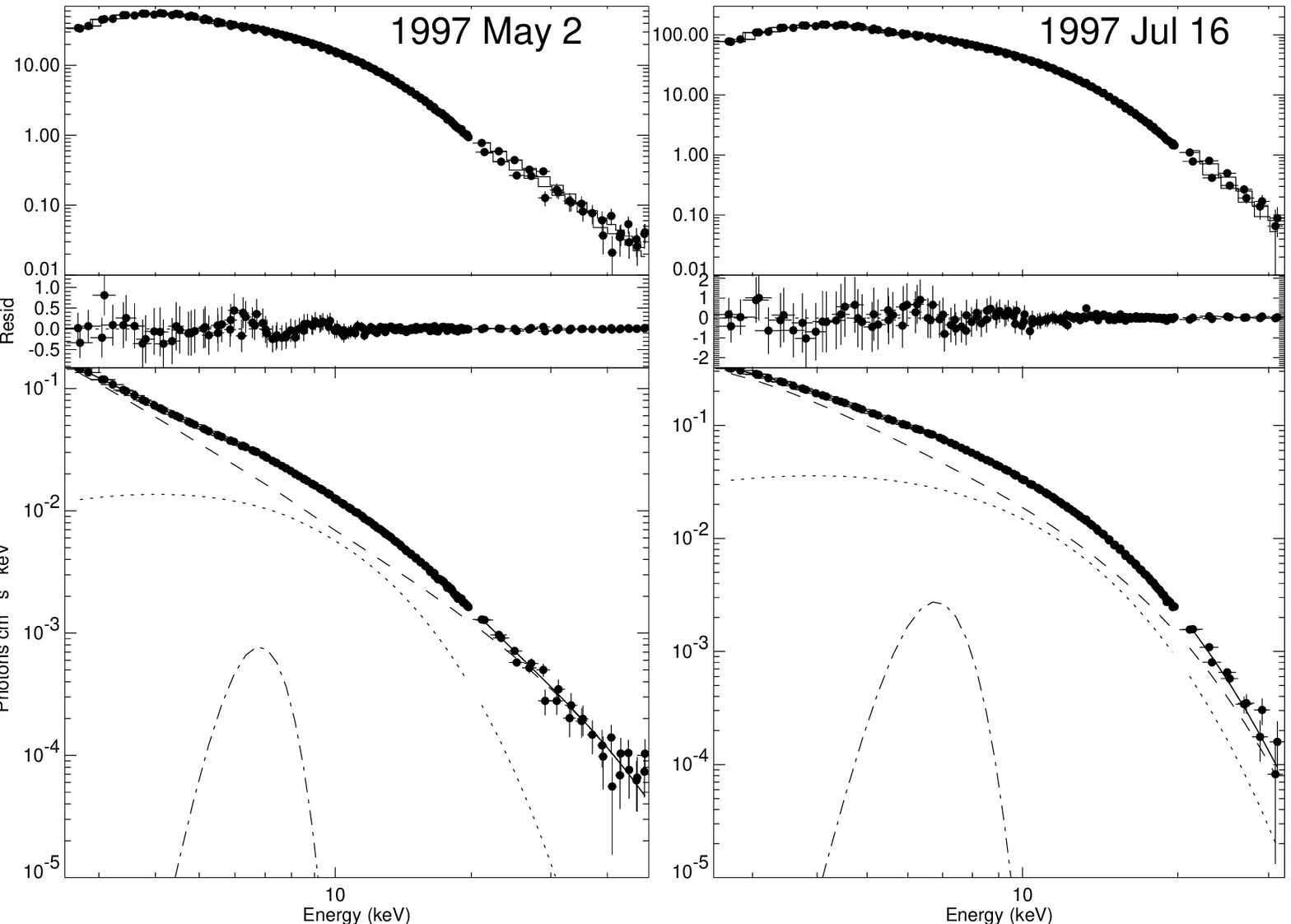]{Examples of 4U 1820-30 spectra fit with the
CompTT + BB 
model.  The top panel shows the raw count rate spectrum with the fitted
model folded through the instruments' responses,  the middle panel shows
the residuals of the fit, and the bottom panel shows the unfolded
spectrum with the individual model components.  The dotted line is the
blackbody, the dashed line is the CompTT model, and the dash-dot line
is the gaussian line.  On the left is the island state spectrum
of 1997 May 2, and on the right is the banana state spectrum of 1997
July 16 (first segment); see Table~\ref{tab-fits2} for the fitted
parameters. 
\label{fig-fit}}

\figcaption[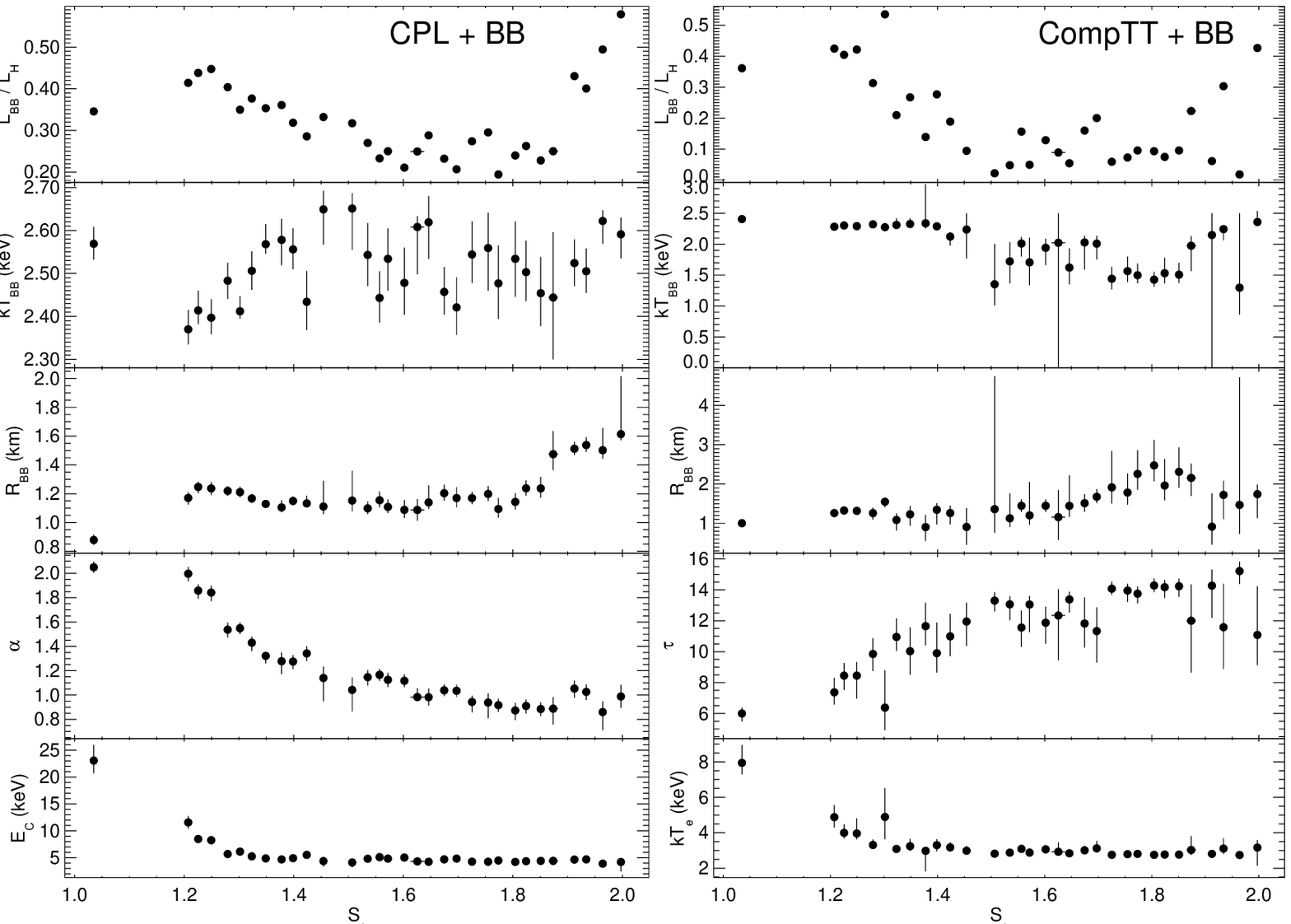]{Variation of spectral parameters
for CPL + BB and 
CompTT + BB models
with accretion rate, 
parameterized by $S_a$.  Here $kT_{BB}$ = blackbody temperature,
$R_{BB}$ = blackbody radius, $\alpha$ = power law photon index, and $E_c$
= power law cut-off energy,
$\tau$ = Comptonization optical depth and
$kT_e$ = Comptonizing electron temperature.  Error bars are $1\sigma$
for one interesting parameter.
\label{fig-params}}

\figcaption[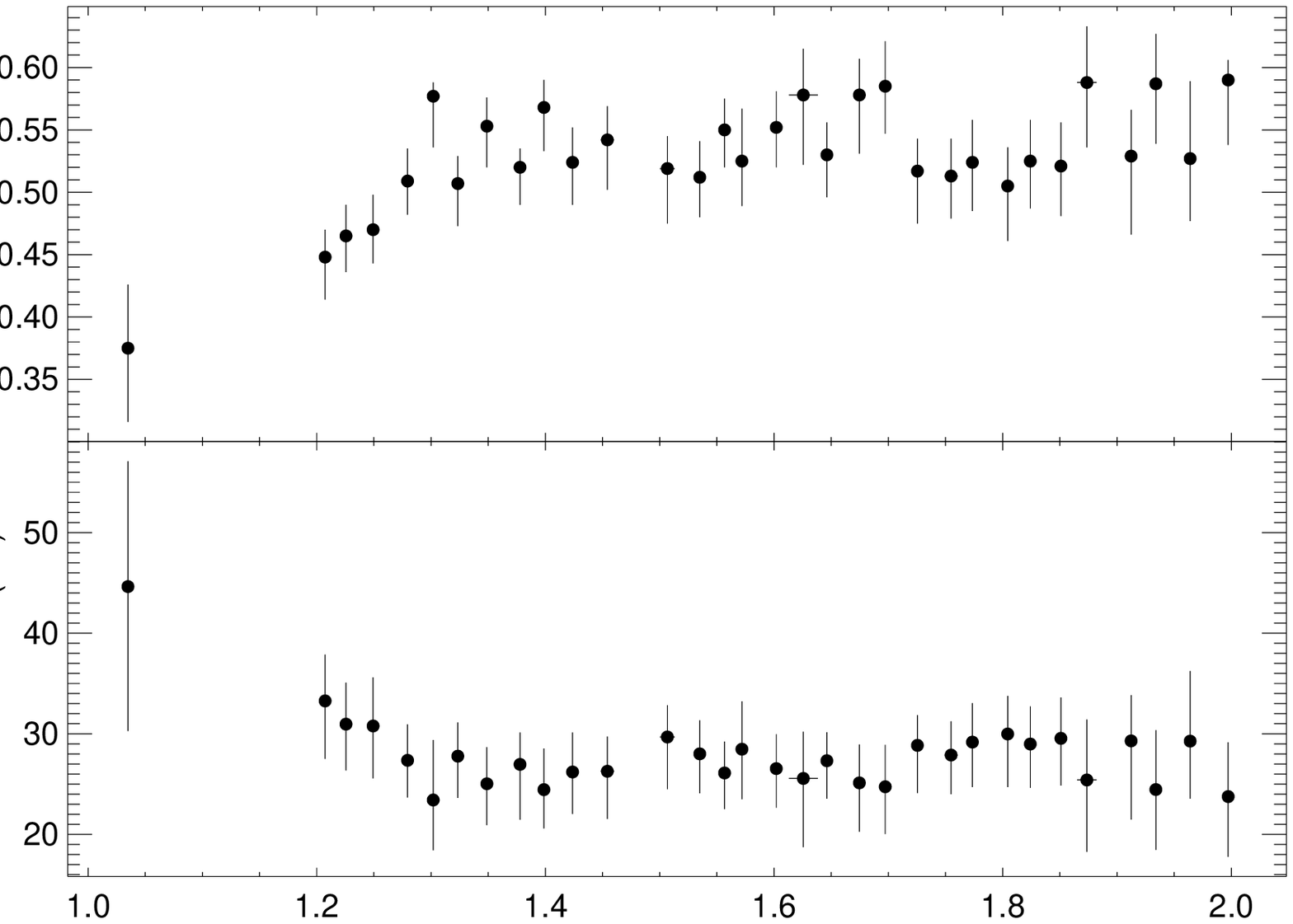]{The temperature of the seed photon
distribution $kT_W$ 
from the CompTT + BB model, with the effective Wien radius $R_W$
described in the text.  The temperature and radius are anticorrelated
below $S_a \sim 1.3$ and constant above.  
\label{fig-wien}}

\figcaption[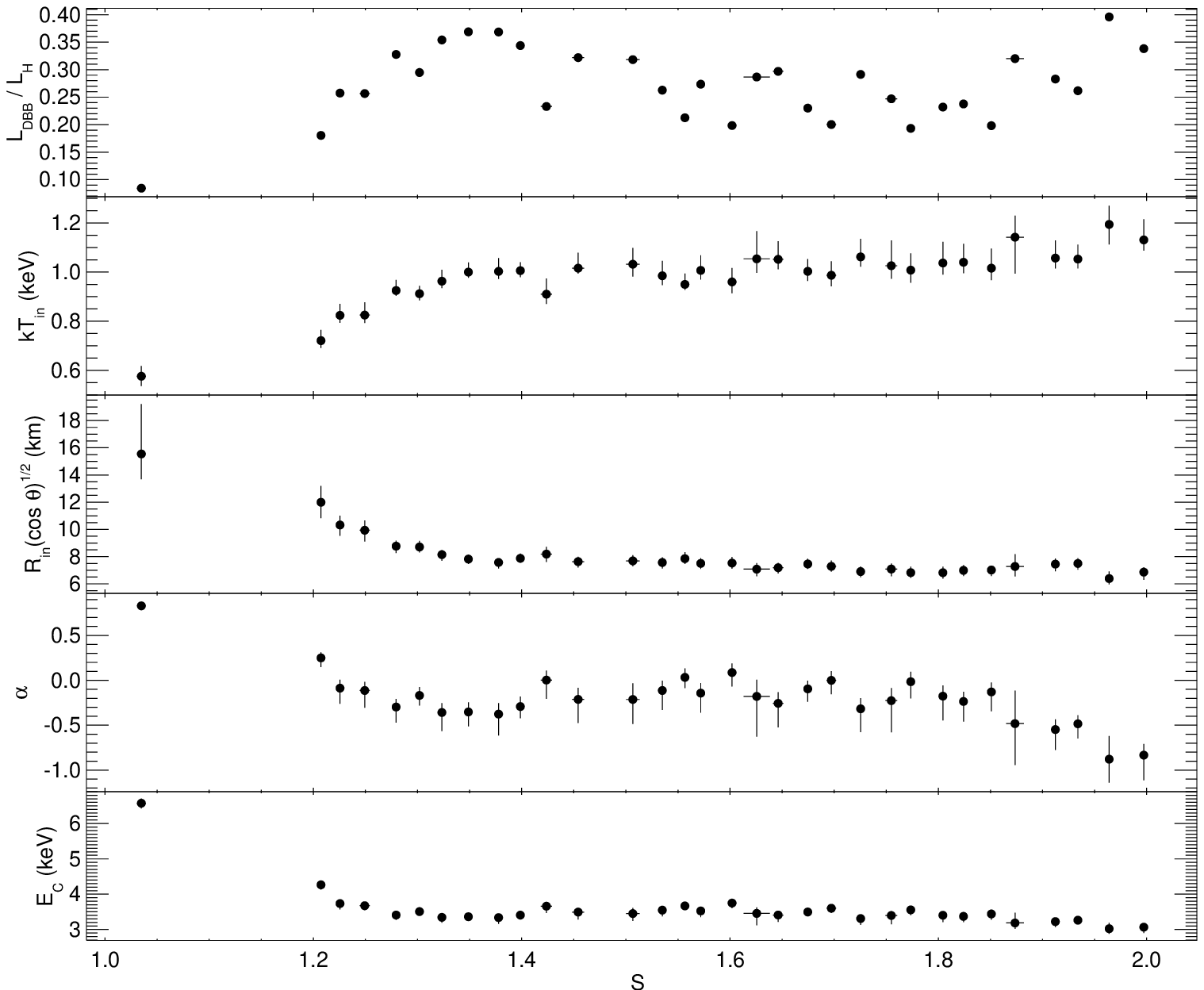]{Variation of  CPL + DBB
parameters with $S_a$.  Here 
$kT_{in}$ = temperature of inner disk and
$R_{in}(\cos\theta)^{1/2}$ = inner disk radius with inclination
$\theta$, while $\alpha$ and $E_c$ are as in
Figure~\ref{fig-params}.  Error bars are $1\sigma$
for one interesting parameter.
\label{fig-paramsdbb}}

\figcaption[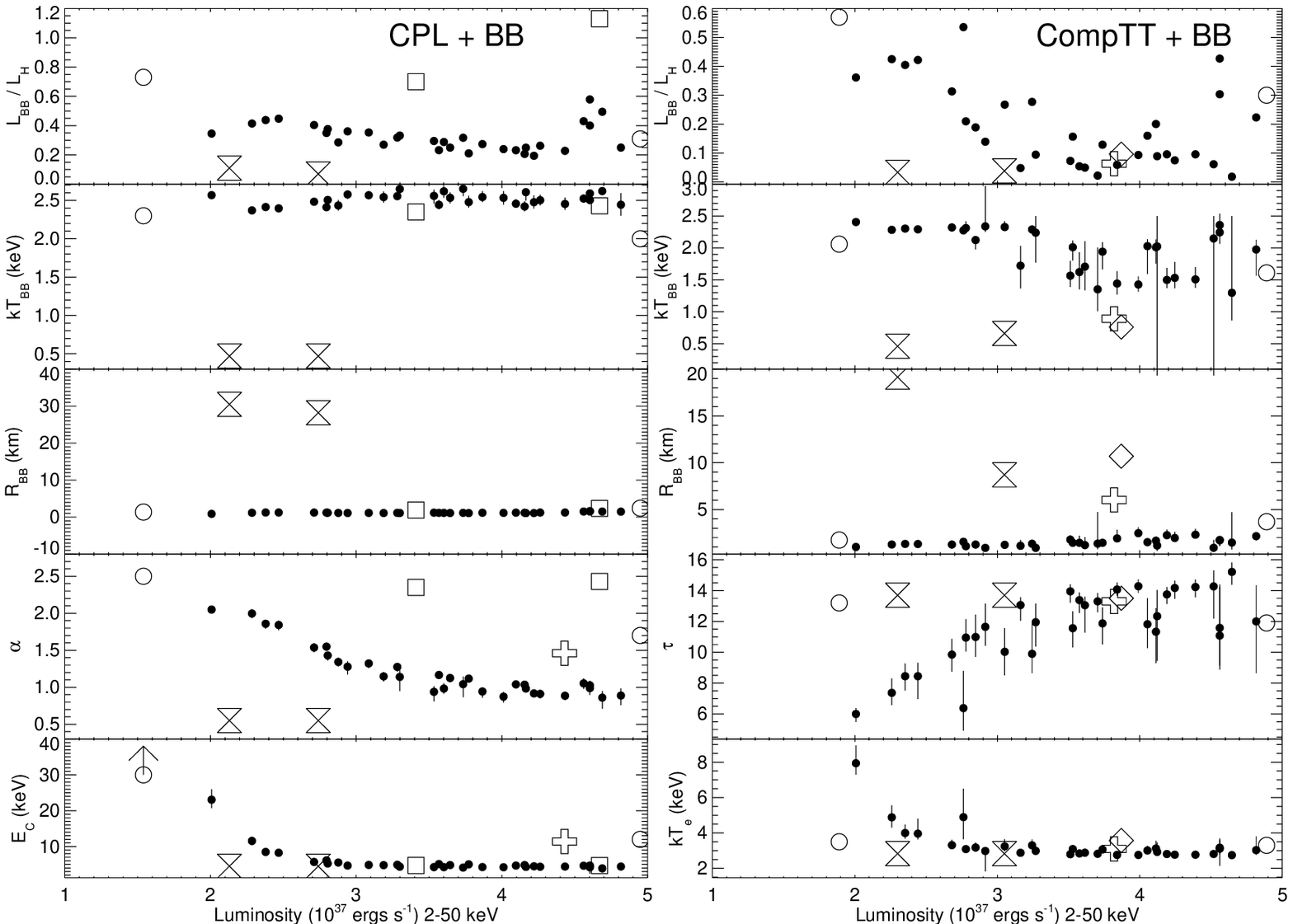]{Comparison of {\em RXTE} spectral
parameters with previous 
measurements (see Table~\ref{tab-prev}) as a function of total 2--50 keV
luminosity.  The luminosities of the previous measurements have
been translated into the 2--50 keV band based on the reported
parameters and fluxes.  {\em EXOSAT} = open circle; {\em Ginga} =
square; {\em ASCA} = diamond; {\em Einstein} = cross; {\em BeppoSAX} =
hourglass.  The luminosities of the {\em RXTE} fits have been
divided by a factor of 1.15 based on fits to the Crab (see text).
\label{fig-comp}}

\figcaption[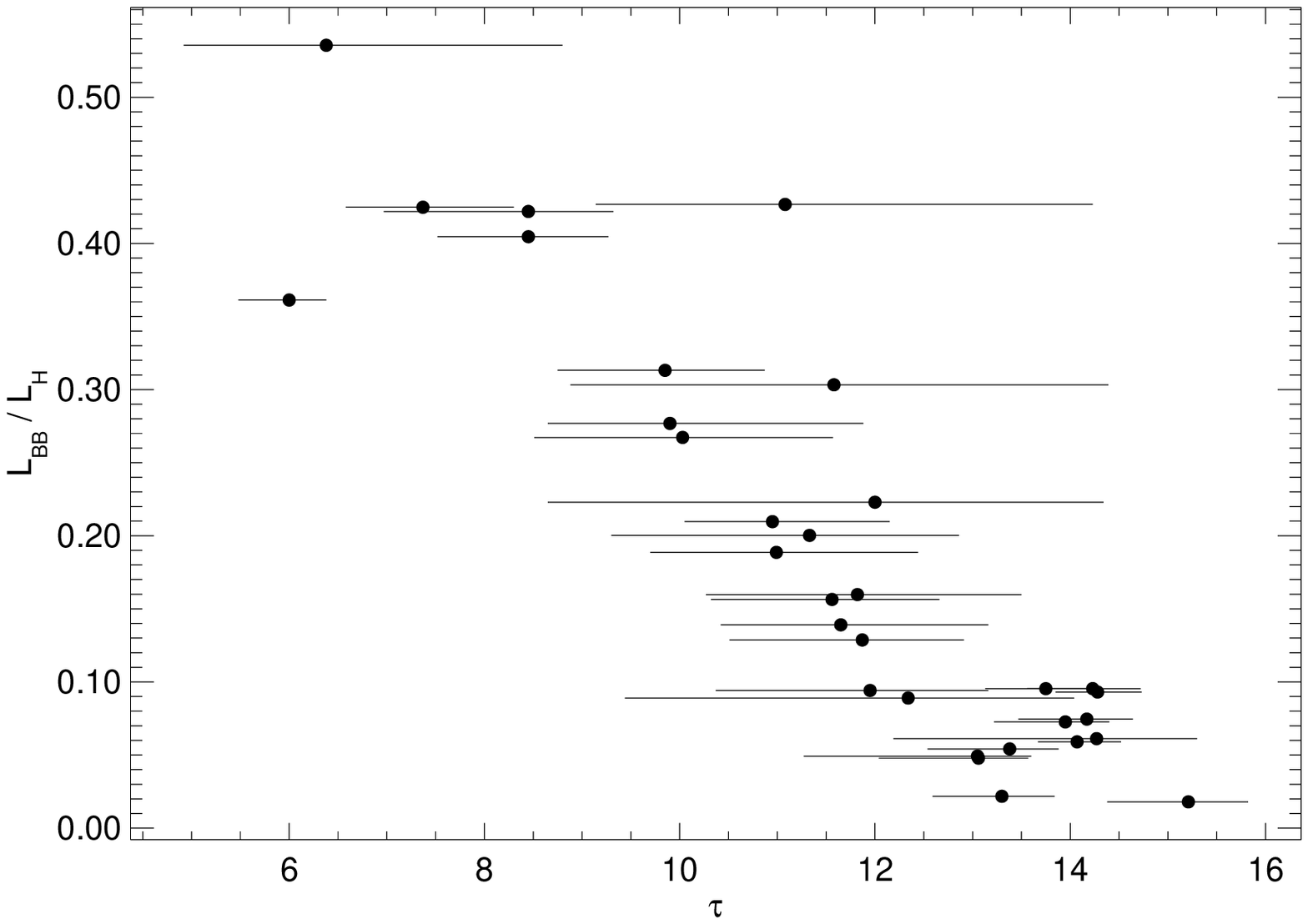]{The ratio of blackbody luminosity to
CompTT luminosity 
$L_{BB}/L_H$ as a function of $\tau$ in the CompTT + BB model.  The
two are strongly anticorrelated; as the optical depth of the
Comptonizing cloud increases, the relative contribution of the
blackbody decreases.
\label{fig-tauratio}}

\clearpage

\begin{deluxetable}{lccccccccccc}
\tiny
\tablecaption{Previously-Reported Spectral Fits of 4U 1820-30\tablenotemark{a}\label{tab-prev}} 
\tablewidth{0pt}
\tablehead{
\colhead{Date(s)} & 
\colhead{$N_H$\tablenotemark{b}} &
\colhead{$kT_{BB}$\tablenotemark{c}} & 
\colhead{$\alpha$\tablenotemark{d}} & 
\colhead{$E_c$\tablenotemark{e}} & 
\colhead{$kT_e$\tablenotemark{f}} & 
\colhead{$\tau$\tablenotemark{g}} & 
\colhead{$E_l$\tablenotemark{h}} & 
\colhead{$L$\tablenotemark{i}} & 
\colhead{$L_{BB}/L_H$\tablenotemark{j}} &
\colhead{$\chi^2_{\nu}$} & 
\colhead{Ref} \\
\colhead{} & 
\colhead{($10^{22}$ cm$^{-2}$)} & 
\colhead{(keV)} &
\colhead{} & 
\colhead{} &
\colhead{(keV)} & 
\colhead{(keV)} &
\colhead{(keV)} & 
\colhead{} &
\colhead{} &
\colhead{} &
\colhead{} 
}
\startdata
\tableline\\[-1.5ex]
\multicolumn{12}{c}{{\em ANS} HXX \hspace{2ex} (1--28 keV)} \\[0.8ex]
\tableline
 1975 Mar 24-8 & $1.3 \pm 0.3$ & ... & $2.12 \pm 0.08$ & ... & ... &
... & $\sim 6.7$ & 14.1\tablenotemark{k} & ... & 1.64 & 1 \\
 1975 Sep 26-30 & $< 1.0$ & ... & $1.41 \pm 0.30$ & ... & ... & ... &
$\sim 6.7$ & 3.2\tablenotemark{k} & ... & 1.03 & 1 \\
 1976 Mar 23-7 & $0.93 \pm 0.3$ & ... & $1.89 \pm 0.08$ & ... & ... &
... & $\sim 6.7$ & 14.9\tablenotemark{k} & ... & 3.6 & 1 \\
\tableline\\[-1.5ex]
\multicolumn{12}{c}{{\em EXOSAT} ME \hspace{2ex} (1--30 keV)} \\[0.8ex]
\tableline
 1984 Sep 26 & $0.4 \pm 0.1$ &  2.0 & 1.7 &
12.0 & ... & ... & ... & 6.0 & 0.25 & ... & 2,3 \\
 1985 Apr 16 & & & & & & & & & & & \\
 1985 Sep 22-3 & & & & & & & & & & & \\
 & 0.39 & 1.61 & ... & ... & 3.3 & 11.9 & 6.7 & 5.9 & 0.25 & 1.02 &
3 \\
 1985 Aug 19-20 & $0.4 \pm 0.1$ & 2.3 & 2.5 & $> 30$ & ... & ... &
... & 2.0 & 0.5 & ... & 2,3 \\
 & 0.52 & 2.06 & ... & ... & 3.5 & 13.2 & 6.7 & 2.1 & 0.49 & 1.71
& 3 \\
\tableline\\[-1.5ex]
\multicolumn{12}{c}{{\em Ginga} LAC \hspace{2ex} (1--20 keV)} \\[0.8ex]
\tableline
 1987 May 1-2 & $0.95 \pm 0.22$ & $2.35 \pm
0.12$ & $2.21 \pm 0.14$ & $4.71 \pm 0.35$ & ... & ... & 6.7 & 5.3 &
0.36 & 0.26 & 4 \\
 1987 May 4-5 & $0.99 \pm 0.25$ & $2.43 \pm 0.08$ & $2.16 \pm
0.16$ & $4.69 \pm 0.38$ & ... & ... & 6.7 & 6.6 & 0.58 & 0.20 & 4 \\
\tableline\\[-1.2ex]
\multicolumn{12}{c}{{\em ASCA} GIS \hspace{2ex} (0.6--11 keV)} \\[0.8ex]
\tableline
 1993 Oct 6 & ... & $0.76 \pm 0.02$ & ... & ... & $3.57 \pm 0.24$ &
$13.5 \pm 0.6$ & 6.7 & 3.7 & 0.15 & 1.63 & 5 \\
\tableline\\[-1.5ex]
\multicolumn{12}{c}{{\em Einstein} SSS+MPC \hspace{2ex} (0.5--20 keV)} \\[0.8ex]
\tableline
1978 Apr 7 & $0.29 \pm 0.002$ & ... & $1.46 \pm 0.03$ & $11.4 \pm 0.2$
& ... & ... & ... & 5.54 & ... & 1.91 & 6 \\
 & 0.27 & 0.89 & ... & ... & 3.1 & $13.3 \pm 0.7$ & ... & 5.15 & 0.06
& 1.31 & 6 \\
\tableline\\[-1.5ex]
\multicolumn{12}{c}{{\em BeppoSAX} LECS+MECS+HPGSPC+PDS \hspace{2ex} (0.3--40 keV)} \\[0.8ex]
\tableline
 1998 Apr 17-18 & 0.1 & 0.47 & 0.55 & 4.5 & ... & ... & ... & 2.9--3.6
& 0.25--0.16 & ... & 7,8 \\
 1998 Sep 19-20 & & & & & & & & & & & \\
 & $0.28 \pm 0.03$ & 0.46--0.66 & ... & ... & $2.83 \pm 0.08$ & $13.7
\pm 0.5$ & ... & 3.7--4.8 & 0.06--0.04 & ... & 7,8 \\
\enddata
\tablenotetext{a}{Only shown are fits using a cut-off power law or
Comptonization ({\em CompST} in XSPEC) model, plus an optional blackbody}
\tablenotetext{b}{Hydrogen column density using cross sections of
Morrison \& McCammon (1983)}
\tablenotetext{c}{Blackbody temperature}
\tablenotetext{d}{Cut-off power law photon index}
\tablenotetext{e}{Cut-off power law cut-off energy}
\tablenotetext{f}{Comptonizing electron temperature}
\tablenotetext{g}{Comptonizing cloud optical depth}
\tablenotetext{h}{Energy of gaussian line feature, if included}
\tablenotetext{i}{Total luminosity, $\times 10^{37}$ ergs 
s$^{-1}$, for a distance of 6.4 kpc in the indicated energy band} 
\tablenotetext{j}{Ratio of blackbody flux to hard component flux
(cut-off power law or Comptonization) in indicated energy band}
\tablenotetext{k}{{\em ANS} counts s$^{-1}$, since no energy flux is given}
\tablerefs{
(1) Parsignault \& Grindlay 1978; (2) Stella, White, \& Priedhorsky
1987; (3) White, Stella, \& Parmar 
1988; (4) Ercan et al. 1993; (5) Smale et al. 1994; (6) Christian \&
Swank 1997; (7) Piraino et al. 1999a; (8) Kaaret et al. 1999
}
\end{deluxetable}


\clearpage

\begin{deluxetable}{lcc}
\tablecaption{{\em RXTE} Observations of 4U 1820-30\label{tab-log}}
\footnotesize
\tablewidth{0pt}
\tablehead{
\colhead{Date} & 
\colhead{Starting Time} &
\colhead{Mean Count Rate} \\
\colhead{} &
\colhead{(MET)\tablenotemark{a}} &
\colhead{Cts s$^{-1}$ PCU$^{-1}$ (2--16 keV)}
}
\startdata
1996 Oct 26 &  88954464 &  434.3 \\
1996 Oct 26 &  88977504 & 388.5 \\
1996 Oct 28 &  89133104 & 481.4 \\
1996 Oct 28 &  89161904 & 498.6 \\
1996 Oct 30 &  89294096 & 655.2 \\
1996 Oct 30 &  89322896 & 624.1 \\
1996 Oct 30 &  89350464 & 551.2 \\
 1997 Feb 9 &  98100800 & 694.6 \\
 1997 Mar 6 & 100306640 & 785.8 \\
 1997 Mar 7 & 100312400 & 813.8 \\
1997 Mar 21 & 101575104 & 540.6 \\
1997 Mar 23 & 101753664 & 623.4 \\
1997 Apr 10 & 103308800 & 491.3 \\
1997 May 2 & 105199312 & 312.0 \\
1997 May 10 & 105887120 & 616.3 \\
1997 May 28 & 107432400 & 887.2 \\
1997 Jun 23 & 109688672 & 855.2 \\
1997 Jun 23 & 109707200 & 851.3 \\
1997 Jul 16 & 111647472 & 819.9 \\
1997 Jul 16 & 111670640 & 698.5 \\
1997 Aug 15 & 114259928 & 752.7 \\
1997 Sep 10 & 116477432 & 543.0 \\
1997 Sep 10 & 116488824 & 562.5 \\
\enddata
\tablenotetext{a}{Mission Elapsed Time = seconds since 1994 Jan 1, 0h0m0s UTC}
\end{deluxetable}


\clearpage

\begin{deluxetable}{lccccccccccc}
\footnotesize
\tablecaption{Spectral Fits of 4U 1820-30 with the CPL + BB
Model\label{tab-fits1}} 
\tablewidth{0pt}
\tablehead{
\colhead{$S_a$} & 
\multicolumn{2}{c}{Int Time (s)} & 
\colhead{$kT_{BB}$} &
\colhead{$R_{BB}$\tablenotemark{a}} &
\colhead{$\alpha$} &
\colhead{$E_c$} &
\colhead{$E_l$} &
\colhead{Eqw\tablenotemark{b}} &
\colhead{$L$\tablenotemark{c}} &
\colhead{$L_{BB}/L_H$} &
\colhead{$\chi^2_{\nu}$} \\
\colhead{} &
\colhead{PCA} &
\colhead{HEXTE} &
\colhead{(keV)} &
\colhead{(km)} &
\colhead{} & 
\colhead{(keV)} &
\colhead{(keV)} &
\colhead{(eV)} &
\colhead{} &
\colhead{} &
\colhead{}
}
\startdata
1.03\tablenotemark{d} &  8481 & 2803 & $ 2.57^{+ 0.04}_{- 0.04}$ & $ 0.88^{+ 0.04}_{- 0.03}$ & $ 2.05^{+ 0.05}_{- 0.04}$ & $23.07^{+ 2.91}_{- 2.36}$ & $ 6.53^{+ 0.12}_{- 0.14}$ &  89 & 2.31 & 0.26 & 0.44 \\
1.21 &  4195 & 1362 & $ 2.37^{+ 0.05}_{- 0.03}$ & $ 1.17^{+ 0.04}_{- 0.05}$ & $ 2.00^{+ 0.06}_{- 0.06}$ & $11.56^{+ 1.20}_{- 1.16}$ & $ 6.82^{+ 0.14}_{- 0.15}$ &  76 & 2.63 & 0.29 & 0.59 \\
1.23 &  5908 & 1874 & $ 2.41^{+ 0.05}_{- 0.03}$ & $ 1.25^{+ 0.04}_{- 0.04}$ & $ 1.86^{+ 0.05}_{- 0.07}$ & $ 8.48^{+ 0.63}_{- 0.76}$ & $ 6.84^{+ 0.16}_{- 0.16}$ &  70 & 2.74 & 0.30 & 0.56 \\
1.25 &  3519 & 1229 & $ 2.40^{+ 0.04}_{- 0.04}$ & $ 1.24^{+ 0.04}_{- 0.05}$ & $ 1.84^{+ 0.06}_{- 0.07}$ & $ 8.25^{+ 0.73}_{- 0.75}$ & $ 6.87^{+ 0.18}_{- 0.17}$ &  57 & 2.84 & 0.31 & 0.69 \\
1.28 &  9654 & 3174 & $ 2.48^{+ 0.04}_{- 0.04}$ & $ 1.22^{+ 0.03}_{- 0.03}$ & $ 1.54^{+ 0.06}_{- 0.06}$ & $ 5.68^{+ 0.44}_{- 0.44}$ & $ 6.93^{+ 0.18}_{- 0.18}$ &  33 & 3.12 & 0.29 & 0.46 \\
1.30 & 13763 & 4268 & $ 2.41^{+ 0.04}_{- 0.02}$ & $ 1.21^{+ 0.04}_{- 0.03}$ & $ 1.55^{+ 0.05}_{- 0.05}$ & $ 6.12^{+ 0.31}_{- 0.30}$ & $ 6.79^{+ 0.18}_{- 0.18}$ &  27 & 3.22 & 0.26 & 0.66 \\
1.32 & 15425 & 5006 & $ 2.51^{+ 0.05}_{- 0.04}$ & $ 1.17^{+ 0.03}_{- 0.03}$ & $ 1.43^{+ 0.05}_{- 0.07}$ & $ 5.25^{+ 0.36}_{- 0.37}$ & $ 6.88^{+ 0.18}_{- 0.17}$ &  45 & 3.23 & 0.27 & 0.43 \\
1.35 &  9938 & 2982 & $ 2.57^{+ 0.05}_{- 0.02}$ & $ 1.13^{+ 0.03}_{- 0.03}$ & $ 1.32^{+ 0.03}_{- 0.06}$ & $ 4.86^{+ 0.38}_{- 0.49}$ & $ 6.83^{+ 0.16}_{- 0.16}$ &  33 & 3.55 & 0.26 & 0.56 \\
1.38 &  6985 & 1268 & $ 2.58^{+ 0.05}_{- 0.06}$ & $ 1.10^{+ 0.05}_{- 0.03}$ & $ 1.28^{+ 0.07}_{- 0.11}$ & $ 4.67^{+ 0.47}_{- 0.65}$ & $ 6.88^{+ 0.15}_{- 0.15}$ &  50 & 3.38 & 0.27 & 0.57 \\
1.40 & 10203 & 3130 & $ 2.56^{+ 0.05}_{- 0.05}$ & $ 1.15^{+ 0.03}_{- 0.03}$ & $ 1.27^{+ 0.05}_{- 0.06}$ & $ 4.90^{+ 0.31}_{- 0.36}$ & $ 6.78^{+ 0.16}_{- 0.15}$ &  30 & 3.77 & 0.24 & 0.45 \\
1.42 &  3333 & 1029 & $ 2.43^{+ 0.07}_{- 0.07}$ & $ 1.13^{+ 0.05}_{- 0.02}$ & $ 1.34^{+ 0.06}_{- 0.06}$ & $ 5.52^{+ 0.37}_{- 0.40}$ & $ 6.85^{+ 0.15}_{- 0.15}$ &  53 & 3.31 & 0.22 & 0.61 \\
1.45 &  2546 &  855 & $ 2.65^{+ 0.04}_{- 0.08}$ & $ 1.11^{+ 0.18}_{- 0.05}$ & $ 1.14^{+ 0.09}_{- 0.19}$ & $ 4.36^{+ 0.30}_{- 1.02}$ & $ 6.90^{+ 0.16}_{- 0.16}$ &  43 & 3.79 & 0.25 & 0.60 \\
1.51 &  1840 &  604 & $ 2.65^{+ 0.04}_{- 0.10}$ & $ 1.15^{+ 0.21}_{- 0.07}$ & $ 1.04^{+ 0.10}_{- 0.18}$ & $ 4.11^{+ 0.62}_{- 0.90}$ & $ 6.85^{+ 0.14}_{- 0.15}$ &  55 & 4.29 & 0.24 & 0.60 \\
1.54 &  6145 & 1912 & $ 2.54^{+ 0.07}_{- 0.07}$ & $ 1.10^{+ 0.05}_{- 0.04}$ & $ 1.15^{+ 0.06}_{- 0.07}$ & $ 4.80^{+ 0.32}_{- 0.40}$ & $ 6.86^{+ 0.16}_{- 0.16}$ &  62 & 3.67 & 0.21 & 0.52 \\
1.56 & 10123 & 3135 & $ 2.44^{+ 0.06}_{- 0.06}$ & $ 1.15^{+ 0.06}_{- 0.05}$ & $ 1.17^{+ 0.05}_{- 0.05}$ & $ 5.10^{+ 0.22}_{- 0.23}$ & $ 6.82^{+ 0.14}_{- 0.14}$ &  50 & 4.10 & 0.19 & 0.39 \\
1.57 &  7844 & 2301 & $ 2.53^{+ 0.07}_{- 0.07}$ & $ 1.11^{+ 0.05}_{- 0.04}$ & $ 1.12^{+ 0.06}_{- 0.06}$ & $ 4.82^{+ 0.29}_{- 0.33}$ & $ 6.80^{+ 0.15}_{- 0.17}$ &  58 & 4.19 & 0.20 & 0.64 \\
1.60 &  5008 & 1638 & $ 2.48^{+ 0.08}_{- 0.07}$ & $ 1.09^{+ 0.07}_{- 0.06}$ & $ 1.12^{+ 0.05}_{- 0.05}$ & $ 5.03^{+ 0.24}_{- 0.27}$ & $ 6.85^{+ 0.15}_{- 0.15}$ &  45 & 4.34 & 0.17 & 0.52 \\
1.63 &   566 &  184 & $ 2.61^{+ 0.02}_{- 0.11}$ & $ 1.09^{+ 0.08}_{- 0.07}$ & $ 0.98^{+ 0.07}_{- 0.01}$ & $ 4.31^{+ 0.43}_{- 0.01}$ & $ 6.82^{+ 0.15}_{- 0.14}$ &  45 & 4.79 & 0.20 & 0.92 \\
1.65 &  5417 & 1668 & $ 2.62^{+ 0.06}_{- 0.09}$ & $ 1.14^{+ 0.12}_{- 0.05}$ & $ 0.98^{+ 0.07}_{- 0.07}$ & $ 4.24^{+ 0.39}_{- 0.70}$ & $ 6.93^{+ 0.19}_{- 0.20}$ &  71 & 4.14 & 0.22 & 0.46 \\
1.67 & 15967 & 5063 & $ 2.46^{+ 0.06}_{- 0.05}$ & $ 1.20^{+ 0.06}_{- 0.05}$ & $ 1.04^{+ 0.05}_{- 0.05}$ & $ 4.68^{+ 0.18}_{- 0.19}$ & $ 6.74^{+ 0.12}_{- 0.13}$ &  51 & 4.71 & 0.19 & 0.55 \\
1.70 &  9511 & 2962 & $ 2.42^{+ 0.07}_{- 0.06}$ & $ 1.17^{+ 0.07}_{- 0.06}$ & $ 1.03^{+ 0.05}_{- 0.05}$ & $ 4.82^{+ 0.19}_{- 0.19}$ & $ 6.75^{+ 0.15}_{- 0.16}$ &  55 & 4.78 & 0.17 & 0.55 \\
1.73 &  9645 & 3200 & $ 2.54^{+ 0.08}_{- 0.07}$ & $ 1.17^{+ 0.04}_{- 0.04}$ & $ 0.94^{+ 0.05}_{- 0.09}$ & $ 4.26^{+ 0.25}_{- 0.43}$ & $ 6.82^{+ 0.15}_{- 0.16}$ &  74 & 4.45 & 0.21 & 0.43 \\
1.75 &  3363 & 1030 & $ 2.56^{+ 0.08}_{- 0.10}$ & $ 1.20^{+ 0.05}_{- 0.05}$ & $ 0.94^{+ 0.07}_{- 0.13}$ & $ 4.24^{+ 0.39}_{- 0.73}$ & $ 6.92^{+ 0.18}_{- 0.24}$ &  77 & 4.06 & 0.23 & 0.55 \\
1.77 &  8527 & 2708 & $ 2.48^{+ 0.09}_{- 0.08}$ & $ 1.09^{+ 0.08}_{- 0.06}$ & $ 0.92^{+ 0.05}_{- 0.05}$ & $ 4.47^{+ 0.20}_{- 0.23}$ & $ 6.77^{+ 0.12}_{- 0.14}$ &  69 & 4.85 & 0.16 & 0.49 \\
1.80 &  5726 & 1812 & $ 2.53^{+ 0.09}_{- 0.09}$ & $ 1.14^{+ 0.06}_{- 0.05}$ & $ 0.87^{+ 0.06}_{- 0.08}$ & $ 4.21^{+ 0.26}_{- 0.40}$ & $ 6.75^{+ 0.15}_{- 0.19}$ &  90 & 4.61 & 0.19 & 0.40 \\
1.82 & 10348 & 3097 & $ 2.50^{+ 0.07}_{- 0.07}$ & $ 1.24^{+ 0.06}_{- 0.05}$ & $ 0.91^{+ 0.05}_{- 0.06}$ & $ 4.35^{+ 0.22}_{- 0.27}$ & $ 6.82^{+ 0.15}_{- 0.17}$ &  60 & 4.90 & 0.21 & 0.42 \\
1.85 &  6401 & 1964 & $ 2.45^{+ 0.08}_{- 0.08}$ & $ 1.24^{+ 0.08}_{- 0.06}$ & $ 0.88^{+ 0.06}_{- 0.06}$ & $ 4.40^{+ 0.20}_{- 0.25}$ & $ 6.76^{+ 0.15}_{- 0.18}$ &  68 & 5.10 & 0.19 & 0.46 \\
1.87\tablenotemark{e} &   466 &  161 & $ 2.44^{+ 0.15}_{- 0.14}$ & $ 1.47^{+ 0.16}_{- 0.11}$ & $ 0.89^{+ 0.09}_{- 0.13}$ & $ 4.41^{+ 0.42}_{- 0.72}$ & $ 6.80^{+ 0.00}_{- 0.00}$ &  47 & 5.54 & 0.20 & 0.66 \\
1.91 &  3327 & 1001 & $ 2.52^{+ 0.06}_{- 0.05}$ & $ 1.51^{+ 0.05}_{- 0.04}$ & $ 1.05^{+ 0.07}_{- 0.07}$ & $ 4.64^{+ 0.33}_{- 0.41}$ & $ 6.78^{+ 0.16}_{- 0.16}$ &  43 & 5.25 & 0.30 & 0.60 \\
1.93 &  3347 & 1023 & $ 2.51^{+ 0.05}_{- 0.05}$ & $ 1.54^{+ 0.05}_{- 0.05}$ & $ 1.03^{+ 0.06}_{- 0.07}$ & $ 4.68^{+ 0.29}_{- 0.34}$ & $ 6.73^{+ 0.15}_{- 0.15}$ &  32 & 5.29 & 0.29 & 0.58 \\
1.96 &  3502 & 1015 & $ 2.62^{+ 0.03}_{- 0.05}$ & $ 1.50^{+ 0.15}_{- 0.06}$ & $ 0.86^{+ 0.09}_{- 0.15}$ & $ 3.88^{+ 0.47}_{- 0.70}$ & $ 6.80^{+ 0.00}_{- 0.00}$ &  51 & 5.39 & 0.33 & 0.49 \\
2.00 &  3510 & 1037 & $ 2.59^{+ 0.04}_{- 0.06}$ & $ 1.61^{+ 0.40}_{- 0.04}$ & $ 0.99^{+ 0.09}_{- 0.09}$ & $ 4.20^{+ 0.73}_{- 1.74}$ & $ 6.89^{+ 0.21}_{- 0.20}$ &  32 & 5.29 & 0.37 & 0.65 \\
\enddata
\tablenotetext{a}{Blackbody radius}
\tablenotetext{b}{Gaussian line equivalent width}
\tablenotetext{c}{Total luminosity, $\times 10^{37}$ ergs 
s$^{-1}$, for a distance of 6.4 kpc, 2--50 keV.  This is the raw
luminosity derived from the fit.}
\tablenotetext{d}{Smeared edge included at $8.90^{+0.42 }_{-0.38}$ keV}
\tablenotetext{e}{Only PCUs 0 and 1 available}
\tablecomments{For all fits $N_H$ is frozen at $0.3 \times 10^{22}$
cm$^{-2}$.  Errors are $1\sigma$ for one interesting parameter.}
\end{deluxetable}


\clearpage

\begin{deluxetable}{lccccccccccc}
\footnotesize
\tablecaption{Spectral Fits of 4U 1820-30 with the CompTT + BB
Model\label{tab-fits2}} 
\tablewidth{0pt}
\tablehead{
\colhead{$S_a$} & 
\colhead{$kT_{BB}$} &
\colhead{$R_{BB}$} &
\colhead{$\tau$} &
\colhead{$kT_e$} &
\colhead{$kT_W$\tablenotemark{a}} &
\colhead{$y$\tablenotemark{b}} &
\colhead{$R_W$\tablenotemark{c}} &
\colhead{Eqw} &
\colhead{$L$\tablenotemark{d}} &
\colhead{$L_{BB}/L_H$} &
\colhead{$\chi^2_{\nu}$} \\
\colhead{} &
\colhead{(keV)} &
\colhead{(km)} &
\colhead{} & 
\colhead{(keV)} &
\colhead{(keV)} &
\colhead{} & 
\colhead{(km)} &
\colhead{(eV)} &
\colhead{} &
\colhead{} &
\colhead{}
}
\startdata
1.03\tablenotemark{e} & $ 2.41^{+ 0.03}_{- 0.04}$ & $ 1.00^{+ 0.03}_{- 0.06}$ & $ 6.00^{+ 0.38}_{- 0.52}$ & $ 7.94^{+ 1.01}_{- 0.65}$ & $ 0.38^{+ 0.05}_{- 0.06}$ & 2.24 & $44.63^{+12.45}_{-14.35}$ &  55 & 2.31 & 0.36 & 0.52 \\
1.21 & $ 2.28^{+ 0.02}_{- 0.02}$ & $ 1.26^{+ 0.04}_{- 0.05}$ & $ 7.37^{+ 0.93}_{- 0.79}$ & $ 4.88^{+ 0.68}_{- 0.58}$ & $ 0.45^{+ 0.02}_{- 0.03}$ & 2.08 & $33.27^{+ 4.60}_{- 5.75}$ &  82 & 2.60 & 0.42 & 0.57 \\
1.23 & $ 2.30^{+ 0.03}_{- 0.02}$ & $ 1.33^{+ 0.07}_{- 0.08}$ & $ 8.45^{+ 0.82}_{- 0.93}$ & $ 3.99^{+ 0.48}_{- 0.32}$ & $ 0.47^{+ 0.03}_{- 0.03}$ & 2.23 & $30.96^{+ 4.13}_{- 4.60}$ &  79 & 2.71 & 0.40 & 0.52 \\
1.25 & $ 2.29^{+ 0.03}_{- 0.03}$ & $ 1.32^{+ 0.09}_{- 0.08}$ & $ 8.45^{+ 0.87}_{- 1.48}$ & $ 3.96^{+ 0.85}_{- 0.33}$ & $ 0.47^{+ 0.03}_{- 0.03}$ & 2.21 & $30.78^{+ 4.83}_{- 5.20}$ &  72 & 2.81 & 0.42 & 0.65 \\
1.28 & $ 2.32^{+ 0.05}_{- 0.04}$ & $ 1.26^{+ 0.14}_{- 0.16}$ & $ 9.85^{+ 1.02}_{- 1.10}$ & $ 3.31^{+ 0.31}_{- 0.22}$ & $ 0.51^{+ 0.03}_{- 0.03}$ & 2.51 & $27.36^{+ 3.57}_{- 3.69}$ &  64 & 3.08 & 0.31 & 0.38 \\
1.30 & $ 2.28^{+ 0.02}_{- 0.03}$ & $ 1.55^{+ 0.03}_{- 0.14}$ & $ 6.38^{+ 2.42}_{- 1.46}$ & $ 4.89^{+ 1.61}_{- 1.25}$ & $ 0.58^{+ 0.01}_{- 0.04}$ & 1.56 & $23.42^{+ 5.97}_{- 5.01}$ &  92 & 3.18 & 0.54 & 0.60 \\
1.32 & $ 2.31^{+ 0.11}_{- 0.05}$ & $ 1.09^{+ 0.17}_{- 0.27}$ & $10.95^{+ 1.20}_{- 0.90}$ & $ 3.09^{+ 0.17}_{- 0.18}$ & $ 0.51^{+ 0.02}_{- 0.03}$ & 2.90 & $27.78^{+ 3.36}_{- 4.14}$ &  68 & 3.20 & 0.21 & 0.37 \\
1.35 & $ 2.33^{+ 0.09}_{- 0.04}$ & $ 1.23^{+ 0.21}_{- 0.29}$ & $10.03^{+ 1.54}_{- 1.52}$ & $ 3.24^{+ 0.41}_{- 0.27}$ & $ 0.55^{+ 0.02}_{- 0.03}$ & 2.55 & $25.03^{+ 3.64}_{- 4.11}$ &  77 & 3.51 & 0.27 & 0.47 \\
1.38 & $ 2.34^{+ 0.63}_{- 0.09}$ & $ 0.91^{+ 0.31}_{- 0.35}$ & $11.65^{+ 1.51}_{- 1.23}$ & $ 2.98^{+ 0.21}_{- 1.16}$ & $ 0.52^{+ 0.02}_{- 0.03}$ & 3.17 & $26.96^{+ 3.16}_{- 5.49}$ &  77 & 3.35 & 0.14 & 0.49 \\
1.40 & $ 2.29^{+ 0.06}_{- 0.04}$ & $ 1.35^{+ 0.17}_{- 0.37}$ & $ 9.90^{+ 1.98}_{- 1.25}$ & $ 3.30^{+ 0.33}_{- 0.34}$ & $ 0.57^{+ 0.02}_{- 0.04}$ & 2.53 & $24.45^{+ 4.08}_{- 3.84}$ &  80 & 3.73 & 0.28 & 0.37 \\
1.42 & $ 2.12^{+ 0.06}_{- 0.15}$ & $ 1.26^{+ 0.19}_{- 0.29}$ & $10.99^{+ 1.45}_{- 1.29}$ & $ 3.17^{+ 0.29}_{- 0.23}$ & $ 0.52^{+ 0.03}_{- 0.03}$ & 3.00 & $26.20^{+ 3.92}_{- 4.17}$ &  80 & 3.27 & 0.19 & 0.55 \\
1.45 & $ 2.24^{+ 0.26}_{- 0.47}$ & $ 0.91^{+ 0.48}_{- 0.45}$ & $11.95^{+ 1.21}_{- 1.58}$ & $ 2.98^{+ 0.26}_{- 0.17}$ & $ 0.54^{+ 0.03}_{- 0.04}$ & 3.34 & $26.28^{+ 3.44}_{- 4.75}$ &  76 & 3.76 & 0.09 & 0.53 \\
1.51 & $ 1.35^{+ 0.65}_{- 0.34}$ & $ 1.36^{+ 3.38}_{- 0.60}$ & $13.30^{+ 0.54}_{- 0.71}$ & $ 2.82^{+ 0.08}_{- 0.04}$ & $ 0.52^{+ 0.03}_{- 0.04}$ & 3.90 & $29.68^{+ 3.14}_{- 5.19}$ &  76 & 4.26 & 0.02 & 0.56 \\
1.54 & $ 1.72^{+ 0.31}_{- 0.36}$ & $ 1.13^{+ 0.63}_{- 0.22}$ & $13.06^{+ 0.51}_{- 1.02}$ & $ 2.88^{+ 0.13}_{- 0.06}$ & $ 0.51^{+ 0.03}_{- 0.03}$ & 3.84 & $28.01^{+ 3.33}_{- 3.91}$ &  71 & 3.64 & 0.05 & 0.47 \\
1.56 & $ 2.01^{+ 0.10}_{- 0.21}$ & $ 1.45^{+ 0.16}_{- 0.16}$ & $11.56^{+ 1.10}_{- 1.24}$ & $ 3.09^{+ 0.22}_{- 0.16}$ & $ 0.55^{+ 0.02}_{- 0.03}$ & 3.23 & $26.11^{+ 3.12}_{- 3.60}$ &  82 & 4.06 & 0.16 & 0.36 \\
1.57 & $ 1.71^{+ 0.40}_{- 0.37}$ & $ 1.20^{+ 0.85}_{- 0.24}$ & $13.05^{+ 0.55}_{- 1.78}$ & $ 2.88^{+ 0.24}_{- 0.06}$ & $ 0.52^{+ 0.04}_{- 0.04}$ & 3.84 & $28.47^{+ 4.75}_{- 4.98}$ &  75 & 4.16 & 0.05 & 0.61 \\
1.60 & $ 1.94^{+ 0.15}_{- 0.28}$ & $ 1.45^{+ 0.17}_{- 0.16}$ & $11.87^{+ 1.04}_{- 1.36}$ & $ 3.06^{+ 0.24}_{- 0.15}$ & $ 0.55^{+ 0.03}_{- 0.03}$ & 3.38 & $26.54^{+ 3.41}_{- 3.90}$ &  75 & 4.30 & 0.13 & 0.48 \\
1.63 & $ 2.02^{+ 0.48}_{- 2.02}$ & $ 1.16^{+ 0.69}_{- 0.58}$ & $12.34^{+ 1.70}_{- 2.90}$ & $ 2.93^{+ 0.54}_{- 0.16}$ & $ 0.58^{+ 0.04}_{- 0.06}$ & 3.49 & $25.56^{+ 4.64}_{- 6.83}$ &  92 & 4.74 & 0.09 & 0.85 \\
1.65 & $ 1.62^{+ 0.31}_{- 0.27}$ & $ 1.45^{+ 0.77}_{- 0.28}$ & $13.38^{+ 0.50}_{- 0.84}$ & $ 2.84^{+ 0.10}_{- 0.05}$ & $ 0.53^{+ 0.03}_{- 0.03}$ & 3.98 & $27.32^{+ 2.83}_{- 3.77}$ &  70 & 4.11 & 0.05 & 0.42 \\
1.67 & $ 2.03^{+ 0.11}_{- 0.44}$ & $ 1.51^{+ 0.23}_{- 0.22}$ & $11.82^{+ 1.68}_{- 1.55}$ & $ 3.01^{+ 0.25}_{- 0.20}$ & $ 0.58^{+ 0.03}_{- 0.05}$ & 3.29 & $25.12^{+ 3.81}_{- 4.85}$ &  94 & 4.66 & 0.16 & 0.57 \\
1.70 & $ 2.01^{+ 0.12}_{- 0.26}$ & $ 1.68^{+ 0.19}_{- 0.16}$ & $11.33^{+ 1.53}_{- 2.03}$ & $ 3.12^{+ 0.43}_{- 0.22}$ & $ 0.58^{+ 0.04}_{- 0.04}$ & 3.14 & $24.74^{+ 4.16}_{- 4.69}$ &  92 & 4.73 & 0.20 & 0.57 \\
1.73 & $ 1.44^{+ 0.19}_{- 0.17}$ & $ 1.92^{+ 0.93}_{- 0.41}$ & $14.07^{+ 0.45}_{- 0.40}$ & $ 2.76^{+ 0.04}_{- 0.03}$ & $ 0.52^{+ 0.03}_{- 0.04}$ & 4.27 & $28.85^{+ 3.00}_{- 4.74}$ &  76 & 4.42 & 0.06 & 0.40 \\
1.75 & $ 1.57^{+ 0.23}_{- 0.18}$ & $ 1.78^{+ 0.48}_{- 0.31}$ & $13.95^{+ 0.45}_{- 0.73}$ & $ 2.80^{+ 0.08}_{- 0.05}$ & $ 0.51^{+ 0.03}_{- 0.03}$ & 4.26 & $27.89^{+ 3.36}_{- 3.89}$ &  62 & 4.04 & 0.07 & 0.51 \\
1.77 & $ 1.50^{+ 0.19}_{- 0.13}$ & $ 2.26^{+ 0.60}_{- 0.40}$ & $13.75^{+ 0.47}_{- 0.62}$ & $ 2.81^{+ 0.06}_{- 0.04}$ & $ 0.52^{+ 0.03}_{- 0.04}$ & 4.16 & $29.17^{+ 3.88}_{- 4.47}$ &  76 & 4.82 & 0.10 & 0.49 \\
1.80 & $ 1.43^{+ 0.13}_{- 0.12}$ & $ 2.47^{+ 0.65}_{- 0.40}$ & $14.28^{+ 0.45}_{- 0.43}$ & $ 2.76^{+ 0.04}_{- 0.04}$ & $ 0.50^{+ 0.03}_{- 0.04}$ & 4.40 & $29.98^{+ 3.76}_{- 5.28}$ &  75 & 4.59 & 0.09 & 0.38 \\
1.82 & $ 1.53^{+ 0.25}_{- 0.16}$ & $ 1.96^{+ 0.67}_{- 0.38}$ & $14.17^{+ 0.47}_{- 0.70}$ & $ 2.77^{+ 0.07}_{- 0.04}$ & $ 0.52^{+ 0.03}_{- 0.04}$ & 4.35 & $28.98^{+ 3.74}_{- 4.36}$ &  66 & 4.88 & 0.07 & 0.40 \\
1.85 & $ 1.51^{+ 0.19}_{- 0.13}$ & $ 2.31^{+ 0.62}_{- 0.40}$ & $14.23^{+ 0.49}_{- 0.67}$ & $ 2.77^{+ 0.06}_{- 0.04}$ & $ 0.52^{+ 0.03}_{- 0.04}$ & 4.39 & $29.55^{+ 4.06}_{- 4.68}$ &  69 & 5.05 & 0.10 & 0.46 \\
1.87\tablenotemark{f} & $ 1.98^{+ 0.15}_{- 0.41}$ & $ 2.16^{+ 0.36}_{- 0.46}$ & $12.00^{+ 2.34}_{- 3.35}$ & $ 3.03^{+ 0.77}_{- 0.27}$ & $ 0.59^{+ 0.05}_{- 0.05}$ & 3.41 & $25.41^{+ 6.01}_{- 7.15}$ &  70 & 5.54 & 0.22 & 0.61 \\
1.91 & $ 2.15^{+ 0.35}_{- 2.15}$ & $ 0.92^{+ 0.84}_{- 0.46}$ & $14.27^{+ 1.03}_{- 2.08}$ & $ 2.81^{+ 0.22}_{- 0.07}$ & $ 0.53^{+ 0.04}_{- 0.06}$ & 4.48 & $29.30^{+ 4.54}_{- 7.81}$ &  66 & 5.20 & 0.06 & 0.58 \\
1.93 & $ 2.24^{+ 0.05}_{- 0.18}$ & $ 1.72^{+ 0.36}_{- 0.62}$ & $11.58^{+ 2.81}_{- 2.70}$ & $ 3.11^{+ 0.58}_{- 0.31}$ & $ 0.59^{+ 0.04}_{- 0.05}$ & 3.26 & $24.46^{+ 5.90}_{- 5.99}$ &  76 & 5.25 & 0.30 & 0.56 \\
1.96 & $ 1.30^{+ 1.20}_{- 0.43}$ & $ 1.47^{+ 3.24}_{- 0.73}$ & $15.21^{+ 0.61}_{- 0.83}$ & $ 2.75^{+ 0.06}_{- 0.03}$ & $ 0.53^{+ 0.06}_{- 0.05}$ & 4.97 & $29.27^{+ 6.96}_{- 5.71}$ &  60 & 5.34 & 0.02 & 0.46 \\
2.00 & $ 2.36^{+ 0.18}_{- 0.04}$ & $ 1.74^{+ 0.25}_{- 0.61}$ & $11.08^{+ 3.15}_{- 1.94}$ & $ 3.16^{+ 0.42}_{- 1.03}$ & $ 0.59^{+ 0.02}_{- 0.05}$ & 3.04 & $23.76^{+ 5.38}_{- 5.98}$ &  72 & 5.25 & 0.43 & 0.58 \\

\enddata
\tablenotetext{a}{Temperature of seed photons (assumed to follow a
Wien law)}
\tablenotetext{b}{Comptonization parameter $y = 4kT_e\tau^2/m_ec^2$}
\tablenotetext{c}{Equivalent Wien radius of seed photons (see text)}
\tablenotetext{d}{Total luminosity, $\times 10^{37}$ ergs 
s$^{-1}$, for a distance of 6.4 kpc, 2--50 keV.  This is the raw
luminosity derived from the fit.}
\tablenotetext{e}{Smeared edge included at
$7.73^{+1.34}_{-0.36}$ keV} 
\tablenotetext{f}{Only PCUs 0 and 1 available}
\tablecomments{For all fits $N_H$ is frozen at $0.3 \times 10^{22}$
cm$^{-2}$.  The gaussian parameters $E_l = 6.8$ keV and $\sigma =
0.8$.  Errors are $1\sigma$ for one interesting 
parameter.} 
\end{deluxetable}


\clearpage

\begin{deluxetable}{lccccccccc}
\footnotesize
\tablecaption{Spectral Fits of 4U 1820-30 with the CPL + DBB
Model\label{tab-fits3}} 
\tablewidth{0pt}
\tablehead{
\colhead{$S_a$} & 
\colhead{$kT_{in}$\tablenotemark{a}} &
\colhead{$R_{in}\sqrt{\cos\theta}$\tablenotemark{b}} &
\colhead{$\alpha$} &
\colhead{$E_c$} &
\colhead{$E_l$} &
\colhead{Eqw} &
\colhead{$L$\tablenotemark{c}} &
\colhead{$L_{DBB}/L_H$} &
\colhead{$\chi^2_{\nu}$} \\
\colhead{} &
\colhead{(keV)} &
\colhead{(km)} &
\colhead{} & 
\colhead{(keV)} &
\colhead{(keV)} &
\colhead{(eV)} &
\colhead{} &
\colhead{} &
\colhead{}
}
\startdata
1.03 & $ 0.58^{+ 0.04}_{- 0.04}$ & $15.55^{+ 3.66}_{- 1.85}$ & $ 0.83^{+ 0.04}_{- 0.05}$ & $ 6.58^{+ 0.13}_{- 0.15}$ & $ 6.86^{+ 0.22}_{- 0.32}$ & 113 & 2.25 & 0.08 & 0.88 \\
1.21 & $ 0.72^{+ 0.04}_{- 0.03}$ & $11.99^{+ 1.20}_{- 1.16}$ & $ 0.25^{+ 0.06}_{- 0.10}$ & $ 4.26^{+ 0.09}_{- 0.13}$ & $ 6.75^{+ 0.14}_{- 0.09}$ &  54 & 2.59 & 0.15 & 0.61 \\
1.23 & $ 0.82^{+ 0.05}_{- 0.03}$ & $10.33^{+ 0.68}_{- 0.78}$ & $-0.09^{+ 0.09}_{- 0.17}$ & $ 3.73^{+ 0.10}_{- 0.16}$ & $ 6.61^{+ 0.16}_{- 0.30}$ &  91 & 2.71 & 0.20 & 0.55 \\
1.25 & $ 0.82^{+ 0.05}_{- 0.03}$ & $ 9.94^{+ 0.72}_{- 0.82}$ & $-0.11^{+ 0.09}_{- 0.19}$ & $ 3.67^{+ 0.10}_{- 0.13}$ & $ 6.64^{+ 0.18}_{- 0.32}$ &  75 & 2.81 & 0.20 & 0.68 \\
1.28 & $ 0.93^{+ 0.04}_{- 0.02}$ & $ 8.77^{+ 0.40}_{- 0.49}$ & $-0.30^{+ 0.09}_{- 0.17}$ & $ 3.41^{+ 0.08}_{- 0.13}$ & $ 6.72^{+ 0.15}_{- 0.22}$ &  65 & 3.10 & 0.25 & 0.40 \\
1.30 & $ 0.91^{+ 0.03}_{- 0.03}$ & $ 8.71^{+ 0.45}_{- 0.40}$ & $-0.17^{+ 0.09}_{- 0.11}$ & $ 3.51^{+ 0.09}_{- 0.10}$ & $ 6.66^{+ 0.16}_{- 0.19}$ &  59 & 3.19 & 0.23 & 0.65 \\
1.32 & $ 0.96^{+ 0.05}_{- 0.03}$ & $ 8.15^{+ 0.34}_{- 0.45}$ & $-0.36^{+ 0.10}_{- 0.21}$ & $ 3.34^{+ 0.08}_{- 0.15}$ & $ 6.61^{+ 0.18}_{- 0.36}$ & 102 & 3.21 & 0.26 & 0.38 \\
1.35 & $ 1.00^{+ 0.04}_{- 0.02}$ & $ 7.82^{+ 0.33}_{- 0.35}$ & $-0.35^{+ 0.11}_{- 0.16}$ & $ 3.36^{+ 0.09}_{- 0.12}$ & $ 6.70^{+ 0.15}_{- 0.18}$ &  75 & 3.53 & 0.27 & 0.48 \\
1.38 & $ 1.00^{+ 0.05}_{- 0.03}$ & $ 7.57^{+ 0.17}_{- 0.44}$ & $-0.38^{+ 0.12}_{- 0.24}$ & $ 3.34^{+ 0.10}_{- 0.18}$ & $ 6.66^{+ 0.16}_{- 0.28}$ & 108 & 3.37 & 0.27 & 0.51 \\
1.40 & $ 1.01^{+ 0.03}_{- 0.03}$ & $ 7.87^{+ 0.35}_{- 0.32}$ & $-0.29^{+ 0.11}_{- 0.13}$ & $ 3.41^{+ 0.09}_{- 0.10}$ & $ 6.69^{+ 0.14}_{- 0.16}$ &  68 & 3.75 & 0.26 & 0.38 \\
1.42 & $ 0.91^{+ 0.06}_{- 0.04}$ & $ 8.19^{+ 0.53}_{- 0.58}$ & $ 0.00^{+ 0.11}_{- 0.21}$ & $ 3.66^{+ 0.11}_{- 0.19}$ & $ 6.73^{+ 0.14}_{- 0.16}$ &  73 & 3.29 & 0.19 & 0.57 \\
1.45 & $ 1.02^{+ 0.06}_{- 0.02}$ & $ 7.63^{+ 0.35}_{- 0.43}$ & $-0.21^{+ 0.13}_{- 0.26}$ & $ 3.49^{+ 0.12}_{- 0.21}$ & $ 6.76^{+ 0.15}_{- 0.19}$ &  75 & 3.78 & 0.24 & 0.54 \\
1.51 & $ 1.03^{+ 0.07}_{- 0.05}$ & $ 7.69^{+ 0.41}_{- 0.40}$ & $-0.21^{+ 0.18}_{- 0.27}$ & $ 3.45^{+ 0.16}_{- 0.21}$ & $ 6.70^{+ 0.15}_{- 0.21}$ &  95 & 4.28 & 0.24 & 0.57 \\
1.54 & $ 0.99^{+ 0.06}_{- 0.04}$ & $ 7.57^{+ 0.38}_{- 0.43}$ & $-0.11^{+ 0.11}_{- 0.22}$ & $ 3.55^{+ 0.10}_{- 0.18}$ & $ 6.68^{+ 0.16}_{- 0.25}$ &  97 & 3.65 & 0.21 & 0.49 \\
1.56 & $ 0.95^{+ 0.04}_{- 0.02}$ & $ 7.85^{+ 0.47}_{- 0.38}$ & $ 0.03^{+ 0.10}_{- 0.12}$ & $ 3.67^{+ 0.10}_{- 0.11}$ & $ 6.73^{+ 0.13}_{- 0.16}$ &  74 & 4.08 & 0.18 & 0.37 \\
1.57 & $ 1.01^{+ 0.06}_{- 0.04}$ & $ 7.50^{+ 0.38}_{- 0.41}$ & $-0.14^{+ 0.11}_{- 0.22}$ & $ 3.53^{+ 0.10}_{- 0.18}$ & $ 6.64^{+ 0.17}_{- 0.26}$ & 107 & 4.18 & 0.21 & 0.60 \\
1.60 & $ 0.96^{+ 0.06}_{- 0.05}$ & $ 7.53^{+ 0.45}_{- 0.41}$ & $ 0.09^{+ 0.10}_{- 0.16}$ & $ 3.75^{+ 0.11}_{- 0.15}$ & $ 6.76^{+ 0.15}_{- 0.17}$ &  66 & 4.32 & 0.17 & 0.49 \\
1.63 & $ 1.05^{+ 0.11}_{- 0.06}$ & $ 7.08^{+ 0.45}_{- 0.52}$ & $-0.18^{+ 0.19}_{- 0.45}$ & $ 3.46^{+ 0.17}_{- 0.33}$ & $ 6.75^{+ 0.15}_{- 0.17}$ &  77 & 4.77 & 0.22 & 0.87 \\
1.65 & $ 1.05^{+ 0.07}_{- 0.04}$ & $ 7.18^{+ 0.36}_{- 0.42}$ & $-0.26^{+ 0.12}_{- 0.27}$ & $ 3.41^{+ 0.10}_{- 0.20}$ & $ 6.70^{+ 0.20}_{- 0.30}$ & 117 & 4.13 & 0.23 & 0.42 \\
1.67 & $ 1.00^{+ 0.05}_{- 0.04}$ & $ 7.46^{+ 0.39}_{- 0.36}$ & $-0.09^{+ 0.09}_{- 0.14}$ & $ 3.50^{+ 0.08}_{- 0.11}$ & $ 6.68^{+ 0.13}_{- 0.17}$ &  80 & 4.69 & 0.19 & 0.54 \\
1.70 & $ 0.99^{+ 0.06}_{- 0.05}$ & $ 7.28^{+ 0.43}_{- 0.39}$ & $-0.00^{+ 0.10}_{- 0.15}$ & $ 3.60^{+ 0.10}_{- 0.13}$ & $ 6.67^{+ 0.15}_{- 0.21}$ &  85 & 4.76 & 0.17 & 0.55 \\
1.73 & $ 1.06^{+ 0.07}_{- 0.04}$ & $ 6.90^{+ 0.37}_{- 0.41}$ & $-0.32^{+ 0.12}_{- 0.26}$ & $ 3.31^{+ 0.09}_{- 0.18}$ & $ 6.64^{+ 0.17}_{- 0.31}$ & 127 & 4.43 & 0.23 & 0.41 \\
1.75 & $ 1.03^{+ 0.10}_{- 0.05}$ & $ 7.09^{+ 0.41}_{- 0.53}$ & $-0.23^{+ 0.14}_{- 0.35}$ & $ 3.40^{+ 0.12}_{- 0.25}$ & $ 6.71^{+ 0.20}_{- 0.40}$ & 106 & 4.06 & 0.20 & 0.53 \\
1.77 & $ 1.01^{+ 0.07}_{- 0.05}$ & $ 6.82^{+ 0.43}_{- 0.37}$ & $-0.01^{+ 0.11}_{- 0.19}$ & $ 3.56^{+ 0.10}_{- 0.15}$ & $ 6.67^{+ 0.14}_{- 0.21}$ &  98 & 4.84 & 0.16 & 0.48 \\
1.80 & $ 1.04^{+ 0.09}_{- 0.05}$ & $ 6.82^{+ 0.45}_{- 0.43}$ & $-0.18^{+ 0.12}_{- 0.27}$ & $ 3.40^{+ 0.10}_{- 0.20}$ & $ 6.59^{+ 0.17}_{- 0.31}$ & 128 & 4.61 & 0.19 & 0.38 \\
1.82 & $ 1.04^{+ 0.08}_{- 0.04}$ & $ 6.99^{+ 0.39}_{- 0.42}$ & $-0.23^{+ 0.11}_{- 0.22}$ & $ 3.37^{+ 0.09}_{- 0.16}$ & $ 6.68^{+ 0.17}_{- 0.31}$ &  95 & 4.90 & 0.19 & 0.40 \\
1.85 & $ 1.02^{+ 0.08}_{- 0.05}$ & $ 7.02^{+ 0.22}_{- 0.44}$ & $-0.13^{+ 0.11}_{- 0.21}$ & $ 3.44^{+ 0.09}_{- 0.16}$ & $ 6.65^{+ 0.16}_{- 0.26}$ &  96 & 5.10 & 0.17 & 0.45 \\
1.87\tablenotemark{d} & $ 1.14^{+ 0.09}_{- 0.15}$ & $ 7.27^{+ 0.91}_{- 0.72}$ & $-0.48^{+ 0.37}_{- 0.46}$ & $ 3.18^{+ 0.29}_{- 0.16}$ & $ 6.94^{+ 0.29}_{- 0.43}$ & 110 & 5.54 & 0.24 & 0.63 \\
1.91 & $ 1.06^{+ 0.07}_{- 0.04}$ & $ 7.45^{+ 0.40}_{- 0.52}$ & $-0.55^{+ 0.11}_{- 0.23}$ & $ 3.22^{+ 0.08}_{- 0.15}$ & $ 6.68^{+ 0.16}_{- 0.23}$ &  74 & 5.25 & 0.22 & 0.58 \\
1.93 & $ 1.05^{+ 0.06}_{- 0.04}$ & $ 7.50^{+ 0.39}_{- 0.45}$ & $-0.48^{+ 0.10}_{- 0.16}$ & $ 3.26^{+ 0.07}_{- 0.12}$ & $ 6.68^{+ 0.14}_{- 0.15}$ &  51 & 5.29 & 0.21 & 0.55 \\
1.96 & $ 1.19^{+ 0.08}_{- 0.08}$ & $ 6.39^{+ 0.52}_{- 0.43}$ & $-0.88^{+ 0.26}_{- 0.26}$ & $ 3.02^{+ 0.16}_{- 0.14}$ & $ 6.69^{+ 0.24}_{- 0.41}$ & 116 & 5.39 & 0.28 & 0.47 \\
2.00 & $ 1.13^{+ 0.08}_{- 0.04}$ & $ 6.86^{+ 0.36}_{- 0.56}$ & $-0.83^{+ 0.12}_{- 0.28}$ & $ 3.07^{+ 0.08}_{- 0.16}$ & $ 6.80^{+ 0.19}_{- 0.24}$ &  65 & 5.29 & 0.25 & 0.60 \\

\enddata
\tablenotetext{a}{Disk inner temperature}
\tablenotetext{b}{Disk inner radius}
\tablenotetext{c}{Total luminosity, $\times 10^{37}$ ergs 
s$^{-1}$, for a distance of 6.4 kpc, 2--50 keV.  This is the raw
luminosity derived from the fit.}
\tablenotetext{d}{Only PCUs 0 and 1 available}
\tablecomments{For all fits $N_H$ is frozen at $0.3 \times 10^{22}$
cm$^{-2}$.  Errors are $1\sigma$ for one interesting parameter.}
\end{deluxetable}


\clearpage

\begin{figure}
\figurenum{\ref{fig-asm}}
\plotone{asm_lc_paper.eps}
\caption{}
\end{figure}

\clearpage
 
\begin{figure}
\figurenum{\ref{fig-ccd}}
\plotone{1820_ccd_paper.eps}
\caption{}
\end{figure}

\clearpage

\begin{figure}
\figurenum{\ref{fig-twospec}}
\plotone{twospec_paper.eps}
\caption{}
\end{figure}

\clearpage

\begin{figure}
\figurenum{\ref{fig-sa}}
\plotone{1820_sa_paper.eps}
\caption{}
\end{figure}

\clearpage

\begin{figure}
\figurenum{\ref{fig-gauss}}
\plotone{gaussian_paper.eps}
\caption{}
\end{figure}

\clearpage

\begin{figure}
\figurenum{\ref{fig-fit}}
\plotone{fit_spec.eps}
\caption{}
\end{figure}

\clearpage

\begin{figure}
\figurenum{\ref{fig-params}}
\plotone{1820_params_paper.eps}
\caption{}
\end{figure}

\clearpage

\begin{figure}
\figurenum{\ref{fig-wien}}
\plotone{wien_fig.eps}
\caption{}
\end{figure}

\clearpage

\begin{figure}
\figurenum{\ref{fig-paramsdbb}}
\plotone{dbb_cpl_params_paper.eps}
\caption{}
\end{figure}

\clearpage

\begin{figure}
\figurenum{\ref{fig-comp}}
\plotone{comp_params.eps}
\caption{}
\end{figure}

\clearpage

\begin{figure}
\figurenum{\ref{fig-tauratio}}
\plotone{tau_ratio_paper.eps}
\caption{}
\end{figure}

\end{document}